%% file: paper.tex
\documentclass[12pt]{article}
\usepackage{epsfig}
\usepackage{amsmath}
\usepackage{hhline}
\usepackage{amssymb}
\usepackage{times}

\newlength{\dinwidth}
\newlength{\dinmargin}
\begin{document}  
\def\be{\begin{equation}}

\def\bea{\begin{eqnarray}}
\def\eea{\end{eqnarray}}
\def\bit{\begin{itemize}}
\def\eit{\end{itemize}}
\def\cm{\,{\rm cm}}
\def\GeV{\,{\rm GeV}}
\def\xbj{x_{\rm Bj}}
\def\ycut{y_{\rm cut}}
\def\kperp{k_\perp}
\def\mjj{M_{\rm jj}}
\def\mjj{M_{\sf jj}}
\def\sumet{\sum E_{T,{\rm \, Breit}}}
\def\avet{\overline{E}_{T}}
\def\avetsq{\overline{E}_T^{\,2}}
\def\etjet{E_{T,{\rm \, jet}}}
\def\etbreit{E_{T,{\rm \, Breit}}}
\def\etaflab{\eta_{\rm forw,\, lab}}
\def\nsubjet{N_{\rm subjet} (y_{\rm cut})}
\def\avnsubjet{\langle N_{\rm subjet} (y_{\rm cut}) \rangle}
\newcommand{\alps}{\alpha_s}
\newcommand{\alpset}{\alpha_s(E_T)}
\newcommand{\alpsmz}{\alpha_s(M_Z)}
\newcommand{\alpsmu}{\alpha_s(\mu_r)}
\newcommand{\ord}{{\cal O}}
\newcommand{\ordalps}{{\cal O}(\alps)}
\newcommand{\cndof}{\chi^2 / {\rm n.d.f.}}

\renewcommand{\floatpagefraction}{0.88}   

%
\begin{titlepage}


\begin{flushleft}
DESY-00-145  \hfill  ISSN 0418-9833 \\
October 2000
\end{flushleft}

\vspace{2.2cm}

\begin{center}
\begin{Large}

{\bf \boldmath
Measurement and QCD Analysis of Jet Cross Sections
in Deep-Inelastic Positron-Proton Collisions at $\sqrt{s}$ 
of 300\,GeV}

\vspace{1.8cm}

H1 Collaboration

\end{Large}
\end{center}

\vspace{1.3cm}

\begin{abstract}
\noindent
Jet production is studied in the Breit frame in 
deep-inelastic positron-proton scattering over a large range 
of four-momentum transfers $5 < Q^2 < 15\,000\GeV^2$
and transverse jet energies 
$7 < E_T < 60\GeV$.
The analysis is based on data corresponding to an integrated 
luminosity of ${\cal L}_{\rm int} \simeq 33\,{\rm pb}^{-1}$
taken in the years 1995--1997 with the H1 detector
at HERA at a center-of-mass 
energy $\sqrt{s} =300\,{\rm GeV}$.
Dijet and inclusive jet cross sections are measured 
multi-differentially using $\kperp$ and angular ordered 
jet algorithms.
The results are compared to the
predictions of perturbative QCD calculations in 
next-to-leading order in the strong coupling constant $\alps$.
QCD fits are performed in which 
$\alps$ and the gluon 
density in the proton are determined separately.
The gluon density is found to be in good agreement 
with results obtained in other analyses using data
from different processes.
The strong coupling constant is determined to be
      $\alpsmz=0.1186 \pm 0.0059$.   
In addition
an analysis of the data in which both $\alps$ and the gluon 
density are determined simultaneously is presented.

\end{abstract}
\vspace{1cm}
\begin{center}
To be submitted to Eur. Phys. J. C  
\end{center}
\end{titlepage}
\begin{flushleft}
  \input{h1auts}
\end{flushleft}
%
\clearpage
\pagenumbering{arabic}

\section{Introduction}

\noindent
Deep-inelastic lepton-proton scattering (DIS) experiments
have played a fundamental role in establishing 
Quantum-Chromodynamics (QCD) as the theory of the strong 
interaction and in the understanding 
of the structure of the proton.
The lepton inclusive DIS cross section is directly sensitive 
to the quark densities in the proton, but 
gives only indirect information
on the gluon content and on the strong coupling constant $\alps$
via scaling violations of the structure functions.
The production rates of events in which the final state contains
more than one hard jet (besides the proton remnant) 
are, however,
observables which are directly sensitive to both $\alps$ and
the gluon density in the proton. 
These multi-jet cross sections can thus be used to test 
the predictions of perturbative QCD (pQCD)
and allow a direct determination of $\alps$ and the gluon 
density~\cite{heraqcd}.

The large center-of-mass energy $\sqrt{s}$ of $300\GeV$ at HERA
allows multi-jet production in DIS to be studied over large regions
of phase space.
In this paper we present comprehensive measurements of jet 
production in the range of four-momentum transfers squared
$5 < Q^2 < 15\,000\GeV^2$.
Using four different jet algorithms we study multi-differential 
distributions of the dijet and the inclusive 
jet cross sections 
to which
the predictions of pQCD in next-to-leading order in 
$\alps$ are compared.
We identify those observables for which 
theoretical predictions have small uncertainties and perform 
QCD analyses of the jet data 
in which
we determine the value of $\alps$.
A consistent determination of the gluon density in the proton, 
together with the quark densities is obtained in a simultaneous fit
which additionally includes data on the inclusive DIS cross section.
In the last step an analysis of the data in which both $\alps$ 
and the gluon density are determined simultaneously is
presented.

\smallskip

This paper is organized as follows.
In section 2 we give a short description of the theoretical 
framework
and motivate the choice of the jet variables 
to be measured.
The experimental environment and details of the measurement 
procedure are described in section 3 and 
the multi-differential jet cross sections are presented
in section 4.
Finally,
in section 5 we introduce the theoretical assumptions
and the methods which are used in the 
QCD analysis and present the results of the QCD fits.
Numerical values of the results are given as tables in
the appendix.

\section{Jet Production in Deep-Inelastic Scattering}

\subsection{Jet Variables and the 
Breit Frame\label{sec:jetvar}}

\begin{figure}
\centering
 \epsfig{file=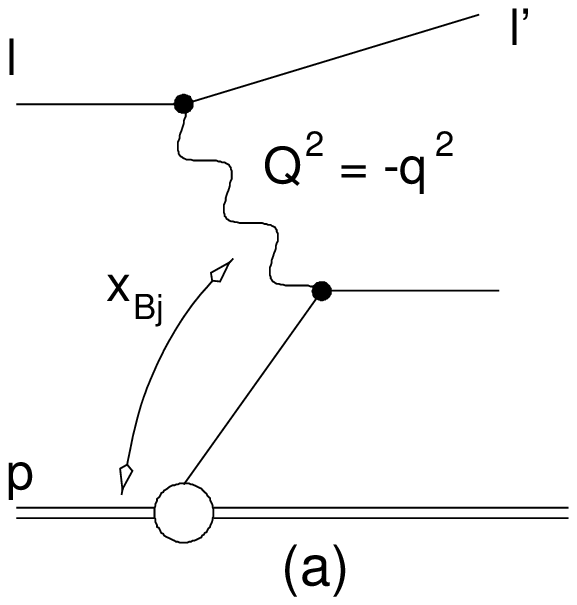,height=4.4cm}\hskip1.7cm
 \epsfig{file=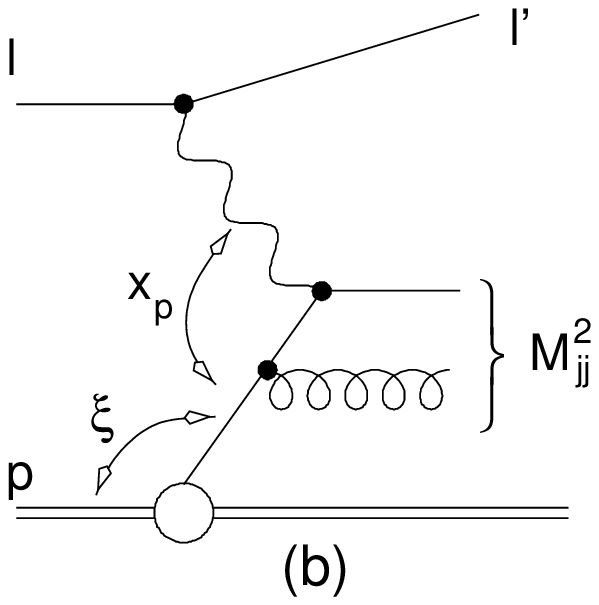,height=4.4cm}\hskip0.5cm
 \epsfig{file=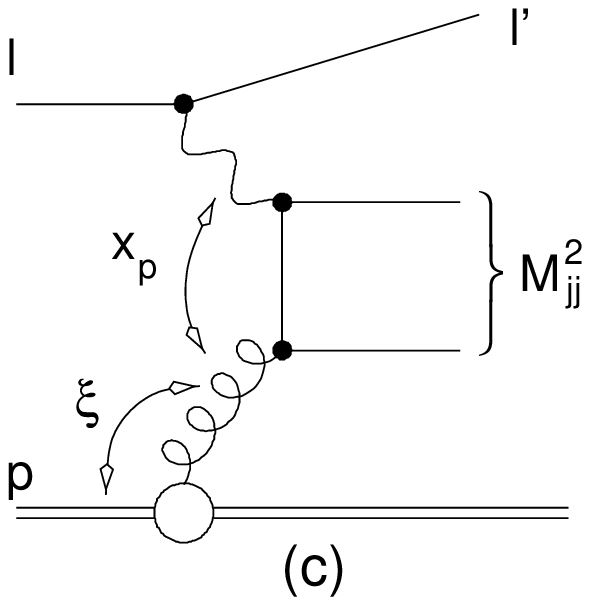,height=4.4cm}
\caption[Diagrams of different processes in deep-inelastic
lepton-proton scattering]%
{Diagrams of different processes in deep-inelastic
lepton-proton scattering: (a) Born process , (b) QCD-Compton 
process and (c) the boson-gluon fusion.}
\label{fig:feynborn}
\end{figure}

The inclusive neutral current cross section in deep-inelastic
lepton-proton scattering is described in lowest order 
perturbation theory as the scattering of the lepton
off a quark in the proton via the exchange of a virtual
gauge boson ($\gamma$, $Z^0$)
(according to Fig.~\ref{fig:feynborn} (a)).
The kinematics of the reaction are given by
the four-momentum transfer squared $Q^2$, the Bjorken scaling 
variable $\xbj$ and the inelasticity $y$ defined as
\be
Q^2 \equiv -q^2 = - (l-l')^2   \hskip11mm
\xbj \equiv \frac{Q^2}{2 p \cdot q}   \hskip11mm
y \equiv \frac{p\cdot q}{p \cdot l}  
\end{equation}
where $l$ ($l'$) and $p$ are the four-momenta of the initial (final)
state lepton and proton, respectively.
When particle masses are neglected
the kinematic variables are related to the lepton-proton
center-of-mass energy $\sqrt{s}$ by $s \, \xbj \, y = Q^2$.
The variable $\xbj$ is in the leading order 
approximation identical with the longitudinal momentum fraction 
$x$ of the proton which is carried by the parton
specified by the parton density functions,
hereafter referred to as the struck 
parton\footnote{In this paper the Bjorken scaling variable 
$\xbj$ is always written with a subscript to distinguish it 
from the proton momentum fraction $x$ which appears in the 
formulae of the proton's parton densities.
While the former is an observable quantity, the latter is only
defined in a theoretical framework within a given factorization 
scheme.}.

Multi-jet production in DIS is described by the
QCD-Compton and the boson-gluon fusion processes.
Due to the latter contribution multi-jet cross sections 
are directly sensitive to the gluon density in the proton.
Examples of leading order diagrams of both processes are shown in 
Fig.~\ref{fig:feynborn} (b) and (c).
Both diagrams contribute to the cross section which 
depends explicitly on $\alps$.
Variables that characterize features of the multi-jet final state
are the invariant mass $\mjj$, the partonic scaling
variable $x_p$ and the variable $\xi$ defined as
\be
\xi \equiv \xbj \, (1+\frac{\mjj^2}{Q^2})  \hskip8mm
\mbox{and}  \hskip8mm
x_p \equiv \frac{\xbj}{\xi} \; .
\label{eq:xidef}
\end{equation}
In the leading order picture (when the final state partons
are identified with jets) the invariant dijet mass $\mjj$ is equal
to the center-of-mass energy of the boson-parton reaction.
In this approximation the fractional momentum $x$ of the struck parton
is given by the variable $\xi$ which becomes much larger than the 
Bjorken scaling variable $\xbj$ if $\mjj$ is large.
The partonic scaling variable $x_p$ specifies the fractional
momentum of the incoming parton seen by the boson.

Studies of the dynamics of multi-jet production are 
preferably performed in the Breit frame
where the 
virtual boson interacts head-on with the proton~\cite{webber93}.
The Breit frame is defined by    
\linebreak[4]
$2\xbj \vec{p} +\vec{q} =0$, where $\vec{p}$ and $\vec{q}$
are the momenta of the proton and the exchanged boson, respectively.
The positive $z$-axis is chosen to be the proton direction.
In the lowest order process, at ${\cal O}(\alps^0)$,
the quark from the proton is back-scattered into the negative 
$z$-direction and no transverse energy is 
produced\footnote{By ``transverse'' we refer to the component
perpendicular to the $z$-axis. The transverse energy is defined
as $E_T \equiv E \sin \theta$.  
The polar angle $\theta$ is defined with respect to the
proton direction in both the laboratory frame and the Breit frame.
Throughout the paper ``transverse energy'' always refers 
to transverse energies in the Breit frame.}.
The appearance of jets with large transverse energies $E_T$
can only be explained by hard QCD processes whose contribution 
is at least of  $\ordalps$ relative to the inclusive DIS 
cross section.
The hardness of the QCD process is specified by $E_T$ 
which is the physical scale at 
which e.g.\  hard gluon radiation is resolved.

In the leading order approximation
the Breit frame is related to the boson-parton
rest frame by a longitudinal boost along the $z$-direction
(see Fig.~\ref{fig:breit_cm}).
The polar scattering angle of the jets in the boson-parton 
center-of-mass frame is directly related to the 
pseudorapidity\footnote{The pseudorapidity is defined
as $\eta \equiv - \ln (\tan \theta / 2)$ where $\theta$
is the polar angle.
Positive values of $\eta$ correspond to particle momenta pointing
into the proton hemisphere.
For massless particles, differences in pseudorapidity
are invariant under longitudinal boosts.}
$\eta'$ of the jets.
In leading order approximation the value of $\eta'$
is equal to half the difference of
the jet pseudorapidities $\eta_{\rm Breit}$ in the Breit frame 
$\eta' = \frac{1}{2} | \eta_{\rm Breit,1} - \eta_{\rm Breit,2}|$.
Since the transverse jet energy $E_T$ is invariant under 
longitudinal boosts along the $z$-axis $E_T$ is identical in 
both frames.

\begin{figure}
\centering
 \epsfig{file=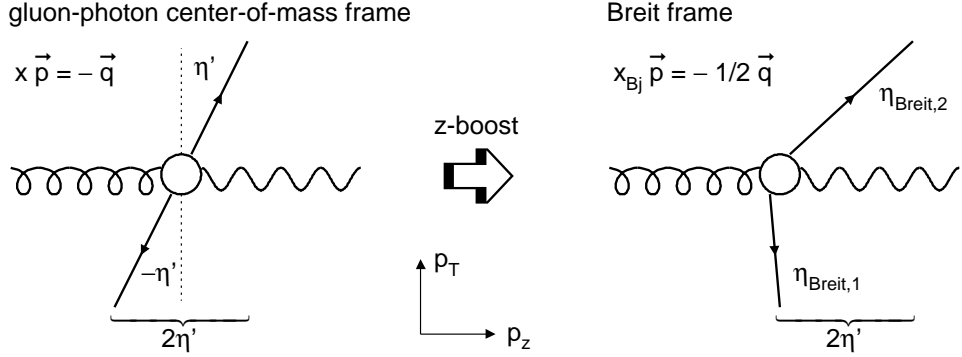,width=13cm}
\vskip-3mm
\caption[A photon-gluon fusion event in different reference frames]%
{A boson-gluon fusion event in deep-inelastic scattering
in the boson-gluon center-of-mass frame (left) and in the Breit frame (right).
The frames are related to each other by a longitudinal boost along 
the $z$-direction.}
\label{fig:breit_cm}
\end{figure}

\subsection{Jet Definitions\label{sec:jetdef}}

The comparison of the properties of high multiplicity 
hadronic final states observed in the experiment to those
in perturbative calculations involving only a small number
of partons requires the definition of infrared- and 
collinear-safe jet observables.
While the properties of different jet observables depend on the 
exact definition of the jets
the physical interpretation of experimental results must, however, 
not depend on details of the jet definition 
if theory is to be claimed successful.

In this analysis we use four jet clustering algorithms which 
successively recombine particles into jets. 
All jet algorithms are applied in the Breit frame to the final state 
particles\footnote{``Particle'' refers in this paper either to an 
energy deposit or a track in the detector, to a parton in a perturbative 
calculation or to a hadron (i.e.\  any particle produced in the 
hadronization process including soft photons and leptons from secondary 
decays). All particles are treated as massless by a redefinition
of the energy ($E\equiv |\vec{p}|$).}
excluding the scattered lepton.
They can be grouped into two pairs of inclusive and exclusive
jet algorithms, each pair consisting of one $\kperp$ ordered and 
one angular ordered algorithm.
In the $\kperp$ (angular) ordered algorithms pairs of particles
are clustered in the order of increasing relative transverse
momenta $\kperp$ (increasing angles) between the particles.
The exclusive jet definitions assign each particle
explicitly to a hard jet or to the proton
remnant, while for the inclusive jet definitions not all 
particles are necessarily assigned to hard jets.
The following four jet algorithms are used:
\begin{itemize} 
\item the exclusive $k_\perp$ ordered algorithm
as proposed in~\cite{ktdis}.

\item the exclusive angular ordered algorithm (Cambridge algorithm)
as proposed in~\cite{cambridge}
and modified for DIS to consider the proton remnant as a particle 
of infinite momentum along the positive $z$-axis, 
following the approach used in \cite{ktdis}. 
The exact definition is taken from~\cite{wobisch2000a}.

\item the inclusive $k_\perp$ ordered algorithm as 
proposed in~\cite{ellissoper,ktclus}.

\item the inclusive angular ordered algorithm (Aachen algorithm)
as proposed in~\cite{wobisch2000a,wobisch99b}.
In analogy to the changes from the exclusive $k_\perp$ algorithm 
to the Cambridge algorithm, the inclusive $k_\perp$ algorithm 
has been modified to obtain an inclusive jet algorithm with 
angular ordering.

\end{itemize}

The recombination of particles in the exclusive jet algorithms 
is made in the $E$-scheme (addition of four-vectors)
resulting in massive jets.
To maintain invariance under longitudinal boosts for the 
inclusive jet definitions 
the $E_T$ recombination scheme~\cite{snowmass} is used
in which the resulting jets are massless. 

In the exclusive jet definitions the clustering procedure is
stopped when the distances $y_{ij} = k_{\perp ij}^2 / S^2$ 
defined between all
pairs of jets and between all jets and the proton remnant
are above some value $\ycut$, where 
$k^2_{\perp ij} = 2 \min(E_i^2,E_j^2) (1-\cos \theta_{ij})$ and
$S$ is a reference scale.
In our analysis we set $S^2=100\GeV^2$ and $\ycut = 1$
to have the final jets separated by 
$k_{\perp ij} > 10\GeV$.
The inclusive jet algorithms are independent of an explicit stopping 
criterion in the clustering procedure and
hard jet selection cuts have to be applied afterwards.
These algorithms are defined by a radius parameter
$R_0$ which we set to $R_0 =1$ as suggested in~\cite{ellissoper}.

\subsection{Phase Space \label{sec:psdef}}

The kinematic region in which the analysis is performed is 
defined by the kinematic variables $y$ and $Q^2$
\be
 0.2 < y < 0.6    \hskip8mm \mbox{and}  \hskip8mm
5 < Q^2 < 15\,000\GeV^2 \;.
\end{equation}
The lower limit on $y$ has been chosen to exclude the kinematic 
region of large $\xbj$ where jets are predominantly produced
in the forward direction, i.e.\  at the edge of the detector 
acceptance.
The upper limit on $y$ ensures large energies of the scattered lepton.

The jet finding is performed using the jet algorithms introduced 
above.
We restrict the jet phase space to the angular range 
in which jets can be well measured in the H1 detector.
Therefore the four-vectors of the jets defined in the Breit frame
are boosted to the laboratory frame where we then apply the 
pseudorapidity cut
\be
-1 < \eta_{\rm jet,\, lab} < 2.5 \;.
\label{eq:etalabcut}
\end{equation}
In this kinematic range double-differential
jet cross sections are measured as a function of 
$Q^2$, $E_T$ (inclusive jet cross section) and
$\avet = \frac{1}{2} (E_{T1} + E_{T2})$ (dijet cross section)
using the various jet algorithms mentioned 
above.
In addition the dependences of the dijet cross section  
on the dijet variables $\mjj$, $\xi$, $x_p$ and $\eta'$ as 
introduced in section~\ref{sec:jetvar} as well as on the 
pseudorapidity of the forward jet $\etaflab$ in the laboratory 
frame are measured.
In all cases inclusive dijet cross sections,
i.e.\   cross sections to produce two or more jets within the 
angular acceptance are measured.
The jet variables are calculated from the two
jets with highest transverse energy (the jets are labeled
in the order of descending $E_T$).

In the measurement of the dijet cross section
care has to be taken to avoid regions at the 
boundary of phase space which are sensitive to soft gluon 
emissions where perturbative calculations in fixed order
are not able to make reliable predictions.
The exclusive jet algorithms avoid these regions due to the 
cut on the variable $\kperp$.
For the inclusive jet definitions additional selection cuts 
have to be chosen appropriately.
The dijet cross section defined by a symmetric cut 
on the transverse energy of the jets
$E_{T,\,1,2} > E_{\rm T, min}$ is infrared sensitive~\cite{frixione97}.
This problem can be avoided
by an additional, substantially harder cut on, for example either 
a) the sum $E_{T,1}+E_{T,2}$,
b) $E_{T,1}$
or 
c) the invariant dijet mass $\mjj$.
When cuts are chosen to obtain cross sections of similar size in 
all of a), b) and c) above, the next-to-leading order corrections 
are largest in b) and hadronization corrections
are largest in c).
The smallest next-to-leading order corrections and 
hadronization effects are seen for scenario a).
For the inclusive jet algorithms
we therefore require
\be
E_{T,\,1,2} > 5\GeV \hskip8mm \mbox{and} \hskip8mm
E_{T,1}+E_{T,2} > 17\GeV \; .
\end{equation}

\subsection{QCD Predictions of Jet Cross Sections\label{sec:theocalc}}

While leading order (LO) calculations can predict the order of 
magnitude and the rough features of an observable, 
reliable quantitative predictions require the perturbative
calculations to be performed (at least) to next-to-leading
order (NLO).
The NLO calculations of the jet cross sections 
used in this analysis are performed in the 
$\overline{\rm MS}$ scheme for five massless quark flavors
using the program 
DISENT\footnote{We have modified DISENT to include the 
running of the electromagnetic coupling constant.}~\cite{disent} 
which has been tested in~\cite{wobisch99c} and found to agree
with the program DISASTER++~\cite{disaster} in the
kinematic region of interest.

Perturbative fixed order calculations beyond leading order
can give reliable quantitative predictions for observables 
with small sensitivity to multiple emission effects and 
non-perturbative contributions.
They fail, however, to predict details of the structure
of multi-particle final states as observed in the experiment.
A complementary approach to describe these properties of the 
hadronic final state is used in parton cascade models.
Starting from the leading order matrix elements, subsequent
emissions are calculated based on soft and collinear 
approximations.
There exist two different approaches in which parton emissions
are either described by a parton shower model
(HERWIG~\cite{HERWIG}, LEPTO~\cite{LEPTO} and 
RAPGAP~\cite{RAPGAP}) or by a dipole cascade 
(ARIADNE~\cite{ARIADNE}).
These parton cascade models can be matched to phenomenological 
models of the hadronization process.
The HERWIG event generator uses the cluster fragmentation
model~\cite{cluster} while in LEPTO, RAPGAP and ARIADNE  
the Lund string model~\cite{lund} is implemented. 
The programs HERWIG, LEPTO, RAPGAP and ARIADNE are
used in the present measurement to provide event samples
which are used in the correction procedure for the data.
In the QCD analysis they are used to estimate the size of the 
hadronization corrections to the perturbative 
jet cross sections.

Higher order QED corrections can change the size of the 
cross section and also modify the event topology.
Especially hard photon radiation may strongly influence
the reconstruction of the event kinematics and thereby 
the boost vector to the Breit frame.
Corrections from real photon emissions from the lepton and 
virtual corrections at the leptonic vertex are included in 
the program HERACLES~\cite{heracles} which is directly 
interfaced to RAPGAP.
An interface to LEPTO and ARIADNE is provided by the program
DJANGO~\cite{django}.

\begin{figure}
\centering
 \epsfig{file=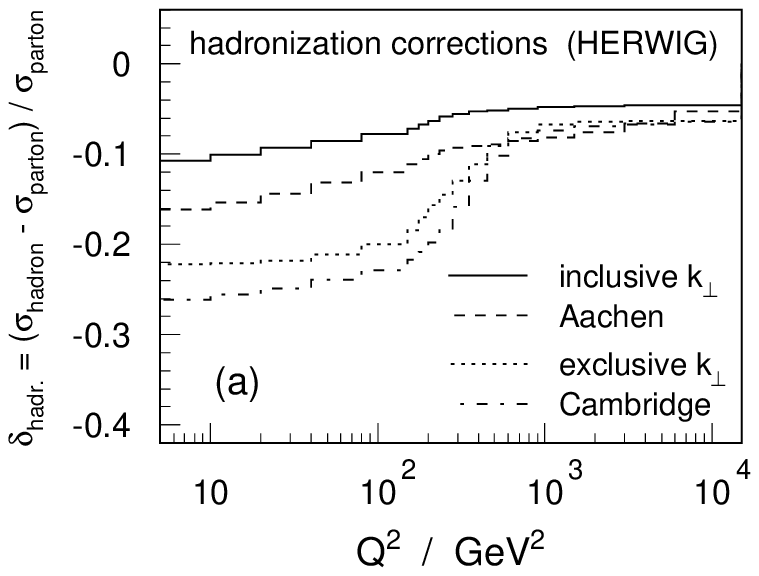,width=7.95cm}
 \epsfig{file=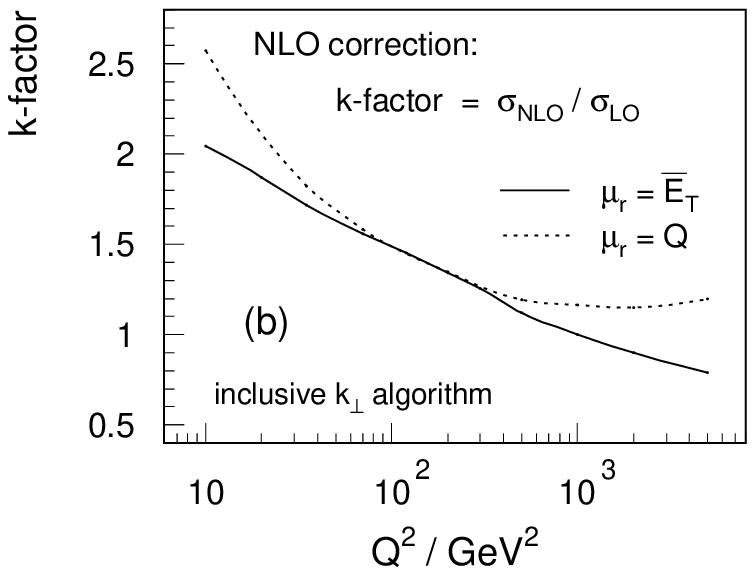,width=7.95cm}
\vskip-3mm
\caption{The predictions of (a) the 
hadronization corrections to the dijet cross section
for different jet definitions as a function of $Q^2$ 
as obtained by HERWIG
and (b) the next-to-leading order corrections to the dijet cross 
section as a function of $Q^2$ for the inclusive $\kperp$ algorithm
using two different renormalization scales $\mu_r$.}
\label{fig:hadcoralgo}
\end{figure}

Safe predictions can only be expected
for observables for which perturbative higher-order 
corrections and non-perturbative (hadronization) 
corrections are small.
Detailed investigations on properties of the NLO cross sections
and the size and the uncertainties of the hadronization corrections 
to the observables under study
have been performed in~\cite{wobisch2000a,wobisch99b,carli99}.
The hadronization corrections predicted from the different
models are in good agreement 
and have small sensitivity to model parameters.
The hadronization corrections $\delta_{\rm hadr.}$ are displayed 
in Fig.~\ref{fig:hadcoralgo} (a) as a function of $Q^2$ for all 
jet algorithms used.
They are defined as $\delta_{\rm hadr.} = (\sigma_{\rm hadron}-
\sigma_{\rm parton}) / \sigma_{\rm parton}$ where
$\sigma_{\rm parton}$ ($\sigma_{\rm hadron}$) is the jet cross 
section before (after) hadronization.
The hadronization corrections are generally smaller for the inclusive 
jet algorithms than for the exclusive ones 
and smaller for the $\kperp$ ordered algorithms
when compared to those with angular ordering.
Hence the inclusive $\kperp$ algorithm shows the
smallest corrections, acceptable even down
to very low $Q^2$ values.
Having the smallest hadronization corrections, the inclusive 
$\kperp$ algorithm is thus the best choice for a jet definition.
The other jet algorithms will, however, still be used
to demonstrate the consistency of the results.

An indication of the possible size of perturbative
higher-order contributions is given by the size of the
NLO corrections or the renormalization and factorization 
scale dependence of an observable.
For the inclusive $\kperp$ algorithm the NLO corrections to the dijet 
cross section are displayed in Fig.~\ref{fig:hadcoralgo} (b).
Shown is the $k$-factor, defined as the ratio of the NLO 
and the LO predictions, for two different choices of the 
renormalization scale ($\mu_r = \avet, Q$).
Towards low $Q^2$ the NLO corrections become large, especially
for the choice $\mu_r=Q$. 
Reasonably small $k$-factors ($k<1.4$) are only seen at
$Q^2 \gtrsim 150\GeV^2$ where $Q^2$ and $E_T^2$ are of similar 
size such that terms $\propto \ln(E_T^2 / Q^2)$ are small.
The renormalization scale dependence is seen to be correlated 
with the NLO correction i.e.\  large at small $Q^2$.
The factorization scale dependence is below 2\% over the whole 
phase space (not shown).
These studies suggest that a QCD analysis of jet cross sections, 
involving the determination of $\alps$ and the gluon density, 
should be performed at large values of $Q^2$.

\section{Experimental Technique}

The analysis is based on data taken in the years 1995--1997
with the H1 detector at HERA in which positrons with energies of
$E_e=27.5\GeV$ collided with protons with energies of
$E_p=820\GeV$.

\subsection{H1 Detector}

A detailed description of the H1 detector can be found elsewhere~\cite{H1det}.
Here we briefly introduce the detector components most relevant 
for this analysis.

In the polar angular range $4^\circ < \theta <  154^\circ$
the electromagnetic and hadronic energy is measured
by the Liquid Argon (LAr) calorimeter~\cite{h1lar} 
with full azimuthal coverage.
The LAr calorimeter consists of an electromagnetic section 
($20-30$ radiation lengths) with lead absorbers and a 
hadronic section with steel absorbers.
The total depth of both sections varies between $4.5$ and $8$ 
interaction lengths. 
Test beam measurements of the LAr calorimeter modules
have shown an energy resolution of 
$\sigma_E/E \approx 0.12/\sqrt{E \;[\GeV]} \oplus 0.01$
for electrons~\cite{h1lar2} and 
$\sigma_{E}/E\approx 0.50/\sqrt{E\;[\GeV]} \oplus 0.02$ for
charged pions after software weighting~\cite{h1pions}.

In the backward direction ($153^\circ < \theta <  177^\circ$)
energy is detected by a lead-fiber calorimeter, SPACAL~\cite{h1spacal}.
It consists of an electromagnetic section with a depth of $28$ 
radiation lengths in which the scattered positron is measured 
with an energy resolution of 
$\sigma_E/E = 0.071/\sqrt{E \;[\GeV]} \oplus 0.010$.
It is complemented by a hadronic section to yield a total depth of 
two interaction length.

Charged particle tracks are measured in two concentric jet drift 
chamber modules (CJC), covering the polar angular range  
$ 25^\circ < \theta < 165^\circ$.  
A forward tracking detector covers $7^\circ < \theta < 25^\circ $ 
and consists of drift chambers with alternating planes 
of parallel wires and others with wires in the 
radial direction. 
A backward drift chamber BDC 
improves the identification of the 
scattered positron in the SPACAL calorimeter.
The calorimeters and the tracking chambers are 
surrounded by a superconducting solenoid providing a uniform
magnetic field of $1.15\,{\rm T}$ parallel to the beam axis in 
the tracking region.

The luminosity is measured using the Bethe-Heitler
process $ep \rightarrow e\gamma p$. 
The final state positron and photon are detected in 
calorimeters situated close to the beam pipe at distances 
of $33\,{\rm m}$ and $103\,{\rm m}$ from the interaction
point in the positron beam direction.

\subsection{Event Selection}

Neutral current DIS events are triggered and identified
by the detection of the scattered positron as a compact 
electromagnetic cluster.
The data set is divided into two subsamples in which the 
positron is detected either in the SPACAL ($5 < Q^2 < 70\GeV^2$) 
or in the LAr calorimeter ($150<Q^2<15\,000\GeV^2$)
with uniform acceptance over the range $0.2 < y < 0.6$.
These regions are labeled ``low $Q^2$'' and ``high $Q^2$''
throughout the text.
The low $Q^2$ and high $Q^2$ samples correspond to
integrated luminosities of 
${\cal L}_{\rm int} \simeq 21\,{\rm pb}^{-1}$ and
${\cal L}_{\rm int} \simeq 33\,{\rm pb}^{-1}$, 
respectively\footnote{The
low $Q^2$ sample uses only data from the years 1996--1997.}.

At low $Q^2$ the positron is reconstructed as the 
highest electromagnetic energy cluster in the SPACAL,
requiring an energy of $E_e' > 10\GeV$ and a polar angle 
of $156^\circ < \theta_{\rm e} < 176^\circ$.
The positron selection at high $Q^2$ closely follows
the procedure used in the recent measurement of the
inclusive DIS cross section~\cite{h1highq2},
requiring an electromagnetic cluster of 
$E_e' >12\GeV$ with a polar angle 
$\theta_{\rm e} \lesssim 153^\circ$.
For $\theta_e > 35^\circ$ the positron candidate is validated 
only if it can be associated with a reconstructed
track, which points to the positron cluster.
Fiducial cuts are applied to avoid the boundary regions
between the calorimeter modules in the $z$ and 
$\phi$ (i.e.\  azimuthal) directions.
The events in the low and high $Q^2$ samples are triggered by 
demanding a localized energy deposition together with loose 
track requirements.
The trigger efficiencies for the final jet event samples are 
above 98\%.

In both samples the reconstructed $z$-coordinate
of the event vertex is required to be within
$\pm 35\cm$ of its nominal position.
The hadronic final state is reconstructed from a combination
of low momentum tracks ($p_T < 2\GeV$) in the central jet 
chamber and energy deposits measured in the LAr calorimeter 
and in the SPACAL 
according to the prescription in~\cite{h1highq2}.
From momentum conservation the sum 
$\sum (E-p_z)$ over all hadronic final 
state particles and the scattered positron is expected
to be $2E_e = 55\GeV$.
This value is lowered in events in which particles
escape undetected in the beam pipe in negative $z$-direction.
Photoproduction background and events with hard photon 
radiation collinear to the positron beam are therefore
suppressed by a cut on $45 < \sum (E-p_z) < 65\GeV$.

The event kinematics is determined from a redundant set of 
variables using the scattered positron and the hadronic 
final state.
Using all hadronic final state particles $h$ the variable 
$\sum = \sum_h (E_h - p_{z,h})$ 
is derived. 
The kinematic variables $\xbj$, $y$ and $Q^2$ are then 
reconstructed by the Electron-Sigma method~\cite{elecsigma}
$$
Q^2_{e\Sigma} = 4 E_e E_e' \cos^2\frac{\theta_e}{2} 
\hskip10mm
x_{e\Sigma} = 
\frac{E_e^{'2} \sin^2 \theta_e}{s \, y_\Sigma (1-y_\Sigma)}
\hskip10mm
y_{e\Sigma} = \frac{2E_e}{\Sigma+E_e'(1-\cos\theta_e)} \: y_\Sigma
$$
\begin{equation}
\mbox{with} \hskip15mm 
y_\Sigma = \frac{\Sigma}{\Sigma+E_e'(1-\cos\theta_e)} \;.
\end{equation}
The jet algorithms are applied to the hadronic final state
particles which are boosted to the Breit frame.
The boost vector is determined from the variables $y_{e\Sigma}$, 
$Q^2_{e\Sigma}$ and the azimuthal angle of the 
scattered positron.
The transverse jet energy $E_T$ (or $\avet$), the dijet mass
$\mjj$ and the variable $\eta'$ are calculated
from the four-vectors of the jets.
The variables $\xi$ and $x_p$ are reconstructed as
\be
\xi_{\rm rec} = x_{\rm e\Sigma} + \frac{\mjj^2}{y_{\rm h} \, s}
\hskip6mm  \mbox{and} \hskip6mm
x_{p, \rm rec} = \frac{x_{e\Sigma}}{\xi_{\rm rec}}
\hskip18mm  \mbox{with} \hskip6mm
y_{\rm h} = \frac{\Sigma}{2E_e}  \; .
\end{equation}
These relations exploit partial cancellations in the 
hadronic energy measurement in $\mjj^2$ and $y_{\rm h}$.

The fraction of dijet events in the inclusive neutral current 
DIS event sample varies strongly with $Q^2$,
namely between 
$\simeq1\%$ (at $Q^2 =5\GeV^2$) and $\simeq20\%$ 
(at $Q^2 = 5000\GeV^2$).
Using the  inclusive $\kperp$ algorithm
we have selected 11400  dijet events at low $Q^2$ 
and 2855 dijet events at high $Q^2$.
The inclusive jet sample (measured only at high $Q^2$)
contains 10\,432 jets with $E_T > 7\GeV$, 
from 7263 events.
The size of photoproduction background has been estimated
using two samples of photoproduction events generated by
PYTHIA~\cite{pythia} and PHOJET~\cite{phojet}.
The contribution to the distributions of the finally selected 
events is found to be negligible (i.e. below 1\%)
in all variables under study.

\subsection{Correction Procedure} 

The data are corrected for effects of limited detector resolution
and acceptance, as well as for inefficiencies of the selection
and higher order QED corrections.
The latter are dominated by real photon emissions from the positron
(initial- and final-state radiation) and virtual
corrections at the leptonic vertex, as included in the program
HERACLES.
No further corrections for effects due to the running of the 
electromagnetic coupling constant or non-perturbative processes
(i.e.\  hadronization) are applied.

To determine the correction functions
the generators LEPTO, RAPGAP and ARIADNE 
(all interfaced to HERACLES) are used.
For each generator two event samples are generated.
The first sample, which includes QED corrections, is subjected to 
a detailed simulation of the H1 detector based on GEANT~\cite{geant}.
The second event sample is generated under the same physics 
assumptions, but without QED corrections and without detector 
simulation.
The correction functions are determined bin-wise for each
observable as the ratio of its value in the second sample and
its value in the first sample.
This method can be used if migrations between different bins
are small and properties of the simulated events are similar
to those of the data.
The absolute normalization of the generated cross sections is, 
however, arbitrary, since this cancels in the ratio.

To test their applicability for the correction procedure,
detailed comparisons have been made of the simulated event 
samples and the data for a
multitude of jet distributions~\cite{wobisch2000a}.
None of the models can describe the magnitude of the
jet cross section; especially at low $Q^2$ large deviations
are seen.
However, all models give a reasonable description
of the properties of the hadronic final state and of the
properties of single jets and the dijet system including
angular jet distributions.
In a previous publication~\cite{h1jetstr} it has 
been shown that these event generators give a good description 
of the internal structure of jets.

Based on the event simulation the bin sizes
of the observables are chosen to match the resolution.
The final bin purities and efficiencies are typically above 
50\%  and migrations are sufficiently small 
to have small correlations between adjacent bins.
The correction functions as determined by different event 
generators are in good agreement with each other and the 
absolute values deviate typically by less than 20\% 
from unity.
The final correction functions applied to the data
are taken to be the average values from the different
models.
The difference between the average and the single values 
are quoted as the uncertainty induced by the model 
dependence which is subdivided in equal fractions into
correlated and uncorrelated uncertainty
between data points.
This uncertainty is typically below 4\%.

\subsection{Experimental Uncertainties\label{sec:expunc}}

In addition to the model dependence of the correction 
function and the statistical uncertainties of the data and of 
the correction function several other sources of 
systematic experimental uncertainties are studied.
They are given in the following, together with the typical
change of the cross sections
and a remark whether a particular uncertainty is treated as 
correlated or uncorrelated between different data points.
The latter classification closely follows the one used
in~\cite{h1highq2}.

\begin{itemize}

\item
The measurement of the integrated luminosity introduces 
an overall normalization uncertainty of $\pm 1.5\%$;
correlated.

\item
The hadronic energy scale of the LAr calorimeter
is varied by $\pm 4\%$; $\pm 2\%$ of the effect
is considered to be correlated; 
typical change of the cross sections $\pm 7.5\%$.

\item
The hadronic energy scale of the SPACAL
is varied by $\pm 7\%$; \\
typical change of the cross sections $< \pm1\%$; uncorrelated.

\item
The track momenta of the hadronic final state
are varied by $\pm 3\%$; \\
typical change of the cross sections $\pm 2.5\%$; uncorrelated.

\item
The calibration of the positron energy in the SPACAL
is varied by $\pm 1\%$;  \\ 
typical change of the cross sections $< \pm 2\%$; correlated.

\item
The positron calibration of the LAr calorimeter 
is treated as in~\cite{h1highq2};
a variation between $\pm 0.7\%$ and $\pm3\%$
is made, depending on the $z$-position of the 
energy cluster in the detector, from which
$\pm 0.5\%$ is considered to be correlated between different 
data points; the rest is treated as uncorrelated;
typical change of the cross sections $\pm 4\%$.

\item
The positron polar angle is varied by $\pm 2\,{\rm mrad}$
($\pm 3\,{\rm mrad}$) for positrons in the SPACAL 
(LAr calorimeter);  \,\,
typical change of the cross sections $< \pm2\%$; correlated.

\item
The positron azimuthal angle is varied by $\pm 3\,{\rm mrad}$; \\
typical change of the cross sections $< \pm 1\%$; uncorrelated.

\end{itemize}
The largest experimental uncertainty comes from the
uncertainty of the energy scale of the LAr calorimeter.
Since the uncertainties from all other
sources are fairly small their determination is often
subject to fluctuations.
We therefore give a conservative estimate, by quoting
the maximal (up- and downward) variation as the symmetric uncertainty.

The statistical and the uncorrelated systematic uncertainties
are added in quadrature to obtain the total uncorrelated
uncertainty.
The correlated contributions are kept separately 
and can thus be considered in a statistical analysis.
To obtain the total uncertainty for each single data point, 
all contributions are added in quadrature.

\section{Experimental Results}

The measured cross sections, corrected for detector effects
and effects of higher order QED, are presented as single-
or double-differential distributions where the inner (outer)
error bars represent the statistical (total)
uncertainty of the data points.
The results (defined in the phase space specified in section
\ref{sec:psdef})
are directly compared to the perturbative QCD 
predictions in NLO.
The hadronization corrections $\delta_{\rm hadr.}$
have been estimated using 
the models described in section~\ref{sec:theocalc} for 
which the predictions are in good agreement with each 
other.
In all the figures shown the theoretical prediction
``NLO $\otimes \, (1 + \delta_{\rm hadr.})$''
is derived from the NLO calculations with hadronization corrections
determined using HERWIG.
All NLO calculations are performed using the parton density
parameterizations CTEQ5M1~\cite{cteq5} and a value of
$\alpsmz =0.118$.
The renormalization scale is set to the transverse jet energy
$\mu_r = E_T$ or in case of the dijet cross section to the 
average transverse energy $\avet$.
For the factorization scale a fixed
value\footnote{This 
slightly unusual procedure is motivated in 
section~\ref{sec:fittec}. 
A variation of $\mu_f$ in the range 
$6 < \mu_f < 30\GeV$ changes the NLO results 
by less than 2\%.}
of $\mu_f = \sqrt{200}\GeV$, corresponding to the average
$E_T$ of the jet sample, is chosen.

\begin{figure}
\centering
 \epsfig{file=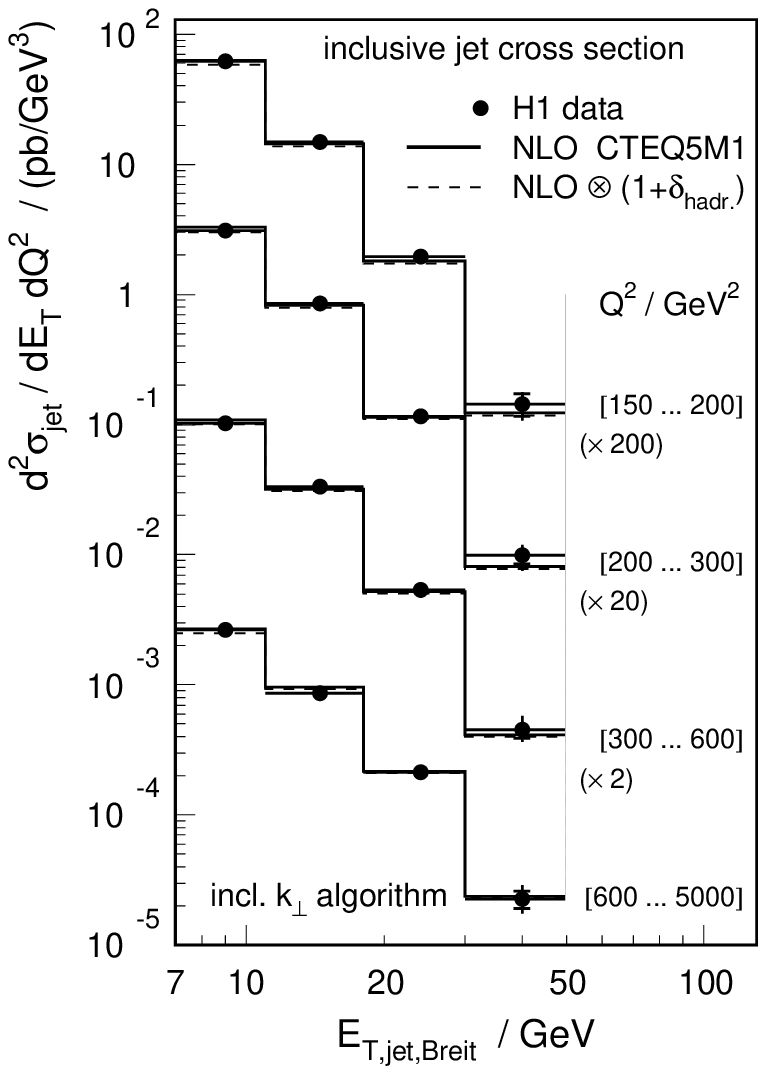,width=7.95cm}
 \epsfig{file=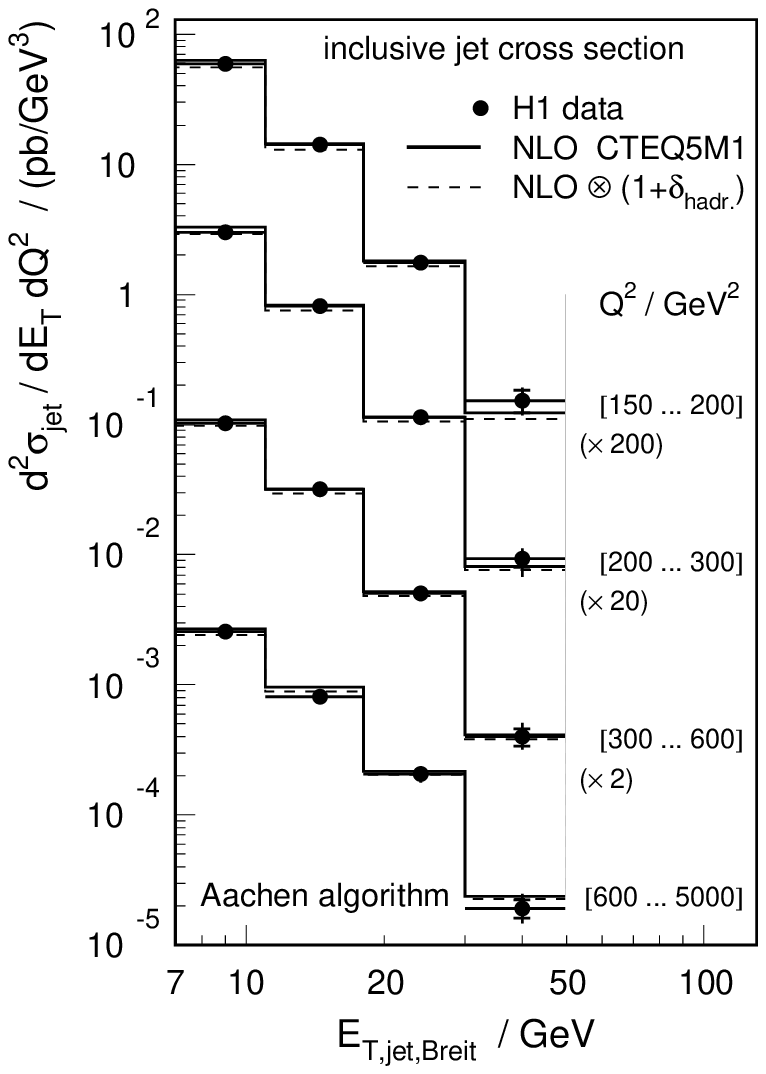,width=7.95cm}
\caption{The inclusive jet cross section as a function of the transverse
jet energy in different regions of $Q^2$ for the inclusive $\kperp$
algorithm (left) and for the Aachen algorithm (right).
The data are compared to the perturbative QCD prediction in NLO 
with (dashed line) and without (solid line) hadronization corrections
included.}
\label{fig:data_incl} 
\end{figure}

\subsection{Inclusive Jet Cross Section}

\begin{figure}
\centering
 \epsfig{file=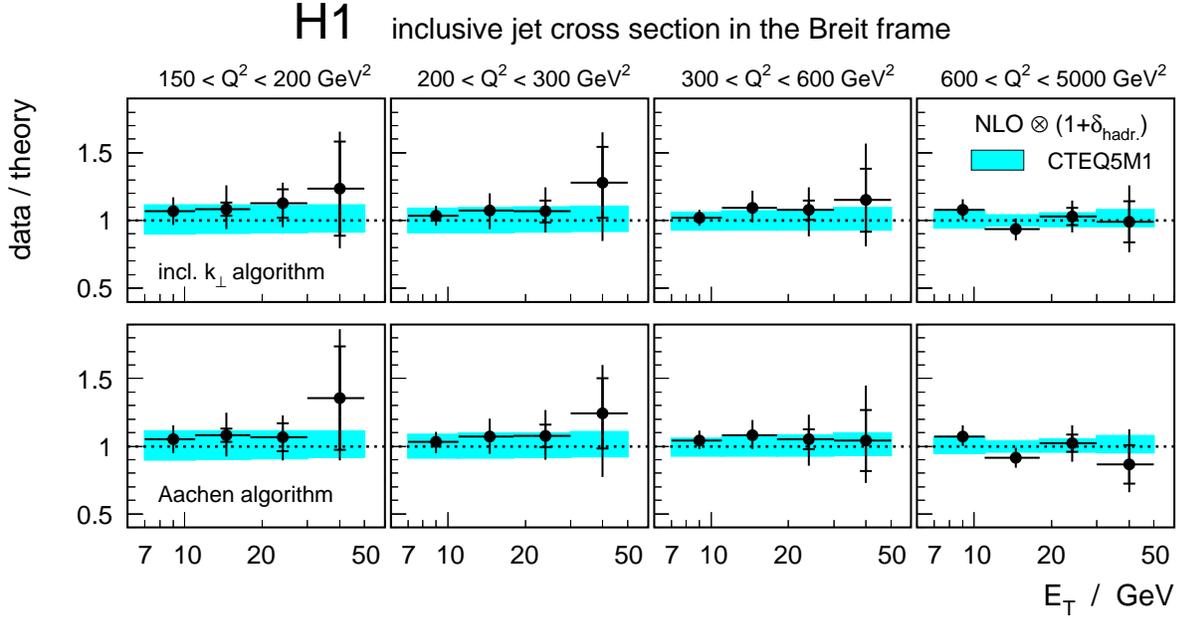}
\caption{The ratio of the measured inclusive jet cross section 
and the theoretical prediction for the inclusive $\kperp$
algorithm (top) and the Aachen algorithm (bottom).
The uncertainty of the theoretical prediction 
is indicated by the band (the contributions from the renormalization 
and factorization scale dependence and the hadronization corrections
are added in quadrature).}
\label{fig:data_incl_ratio} 
\end{figure}

The inclusive jet cross section is measured at high $Q^2$
in the Breit frame for both inclusive jet algorithms.
The results are presented in Fig.~\ref{fig:data_incl} 
double-differentially as a function of the transverse jet energy in 
the Breit frame $E_T$ in different regions of $Q^2$.
The data for the inclusive $\kperp$ algorithm (left) and for the 
Aachen algorithm (right) cover a range of  transverse 
jet energies squared ($ 49 < E_T^2 < 2500 \GeV^2$) which is similar 
to the range of the four-momentum transfers squared 
($150 < Q^2 < 5000\GeV^2$) 
of the event sample.
The cross sections are of the same size  for both jet algorithms 
and show a slightly harder $E_T$ spectrum with increasing $Q^2$.
The hadronization corrections are seen to be below 10\%
for both algorithms.
The ratio of data and theoretical prediction is shown in
Fig.~\ref{fig:data_incl_ratio}.
Over the whole range of $E_T$ and $Q^2$ the NLO calculation,
corrected for hadronization effects, gives a good description
of the data.

\subsection{Dijet Cross Section}

The inclusive dijet cross section is measured over the large range
of four-momentum transfers squared $5< Q^2 < 15\,000\GeV^2$
using the inclusive $\kperp$ jet algorithm.
At high $Q^2$ additional measurements have been made
using the three other jet algorithms.


\begin{figure}
\centering
 \epsfig{file=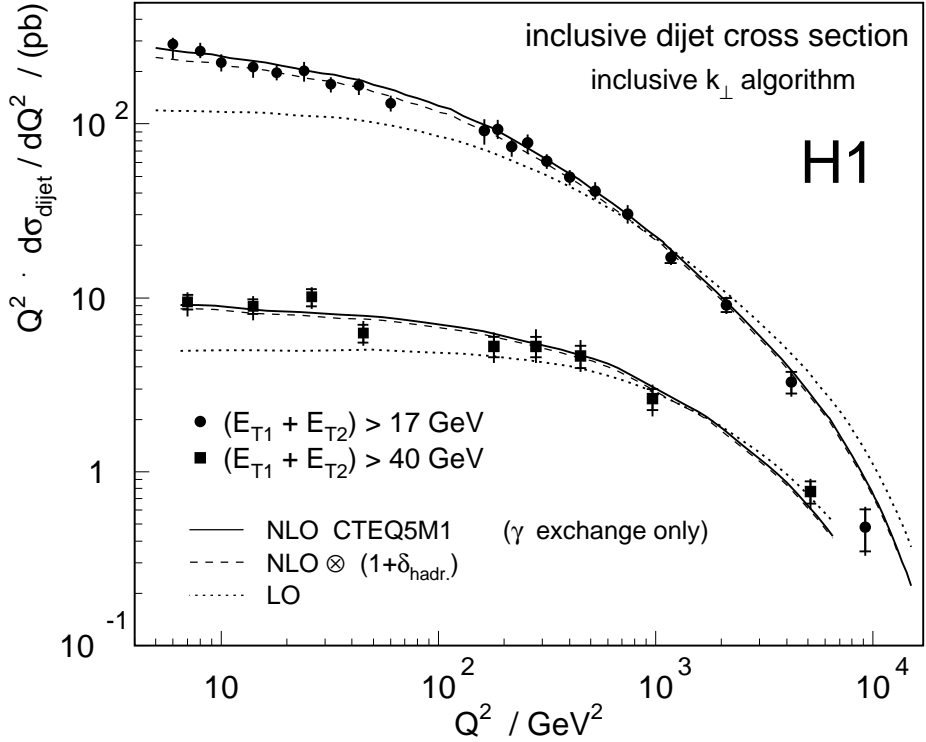,width=12.5cm}
\caption{The dijet cross section measured with the inclusive $\kperp$ 
algorithm as a function of $Q^2$ for different cuts on the
sum of the transverse jet energies.
The data are compared to the perturbative QCD prediction in NLO 
(solid line), in LO (dotted line)
and to a theoretical prediction where hadronization corrections
are added to the NLO prediction (dashed line).}
\label{fig:data_dij_q2tot}
\end{figure}

The dijet cross section for the inclusive $\kperp$ algorithm
is shown in Fig.~\ref{fig:data_dij_q2tot}
as a function of $Q^2$ in the range $5<Q^2<15\,000\GeV^2$
for the central analysis cut ($E_{T1}+E_{T2} > 17\GeV$) and an 
additional harder cut ($E_{T1}+E_{T2} > 40\GeV$) on the sum of 
the transverse energies of the two jets with highest $E_T$.
The data\footnote{All cross sections have been measured as 
bin-averaged cross sections and all but two are presented
this way, the only exception being the presentation 
of the $Q^2$ dependence in
Figs.~\ref{fig:data_dij_q2tot} and~\ref{fig:data_dij_q2all}.
Here, to compare the data 
to the differential NLO prediction the points
are presented at the bin-center, as determined
using the NLO calculation.}
are compared to the NLO prediction without
and with hadronization corrections applied, 
as well as to a LO calculation.

The hadronization corrections are small and increase only 
slightly towards lower $Q^2$.
The NLO prediction, including hadronization effects, 
gives a good description of the data over the large 
phase space in $E_T$ and $Q^2$ and nicely models the 
reduced $Q^2$ dependence observed for the 
higher $E_T$ data. 
Deviations at $Q^2 \simeq 10\,000\GeV^2$ can be attributed
to the neglect of $Z^\circ$ exchange in the calculation.

\begin{figure}
\centering
 \epsfig{file=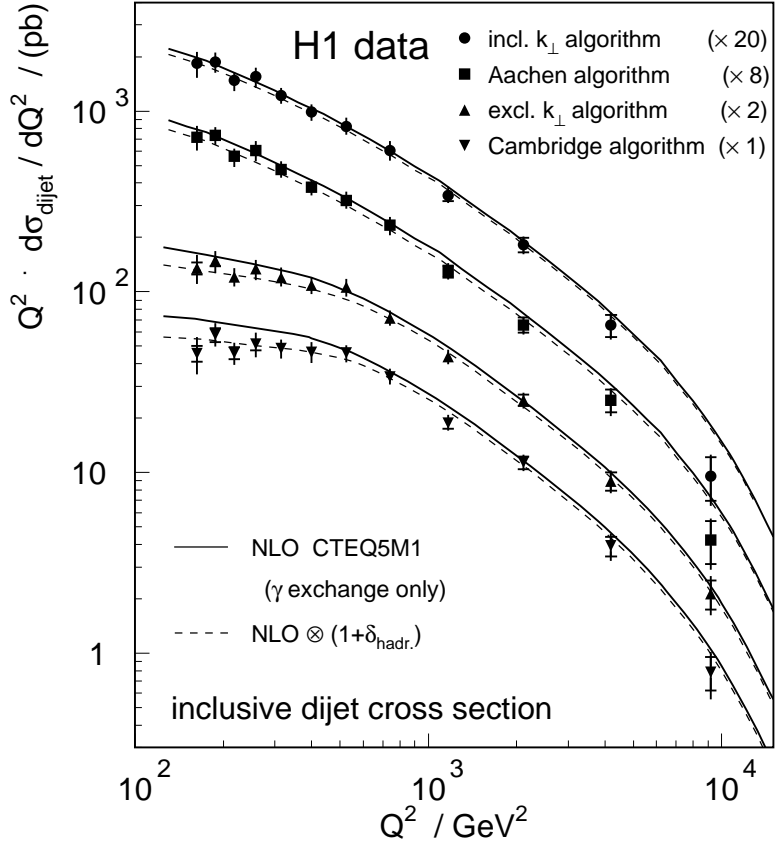,width=10.5cm}
\caption{The dijet cross section as a function of $Q^2$
for four jet algorithms.
The data are compared to the perturbative QCD prediction in NLO 
with (dashed line) and without (solid line) hadronization corrections
included.}
\label{fig:data_dij_q2all}
\end{figure}

The dijet cross section at high $Q^2$, measured using all
four jet algorithms mentioned above,
is shown in Fig.~\ref{fig:data_dij_q2all}.
A different $Q^2$ dependence is observed for the inclusive
and the exclusive algorithms which is a reflection of the 
different jet selection criteria and which is well reproduced by 
the theory.
While hadronization corrections have only a small effect
for the inclusive jet algorithms, they lower the NLO predictions
for the exclusive algorithms by up to 30\% at $Q^2 = 150\GeV^2$.
However, when these non-perturbative corrections are included
the dijet cross sections are in all four cases well described 
by the theoretical curves.

In the following more details of the dijet distributions are given.
For these studies we restrict the phase space to
$Q^2 < 5\,000\GeV^2$ in order to avoid the region 
where contributions from $Z^\circ$ exchange
are sizable.
We present the double-differential dijet cross section 
as a function of the variables $Q^2$, $\mjj$, $\avet$, $\eta'$, 
$x_p$, $\xi$ and $\etaflab$
in Figs.~\ref{fig:data_dij_etmjj} -- \ref{fig:data_dij_etalab}.
As for the $Q^2$ distribution in Fig.~\ref{fig:data_dij_q2all},
the results are compared to the perturbative QCD prediction
in NLO with and without hadronization corrections included.
In Fig.~\ref{fig:data_dij_xpxi} the contribution from gluon-induced
processes is shown in addition.

\begin{figure}
\centering
 \epsfig{file=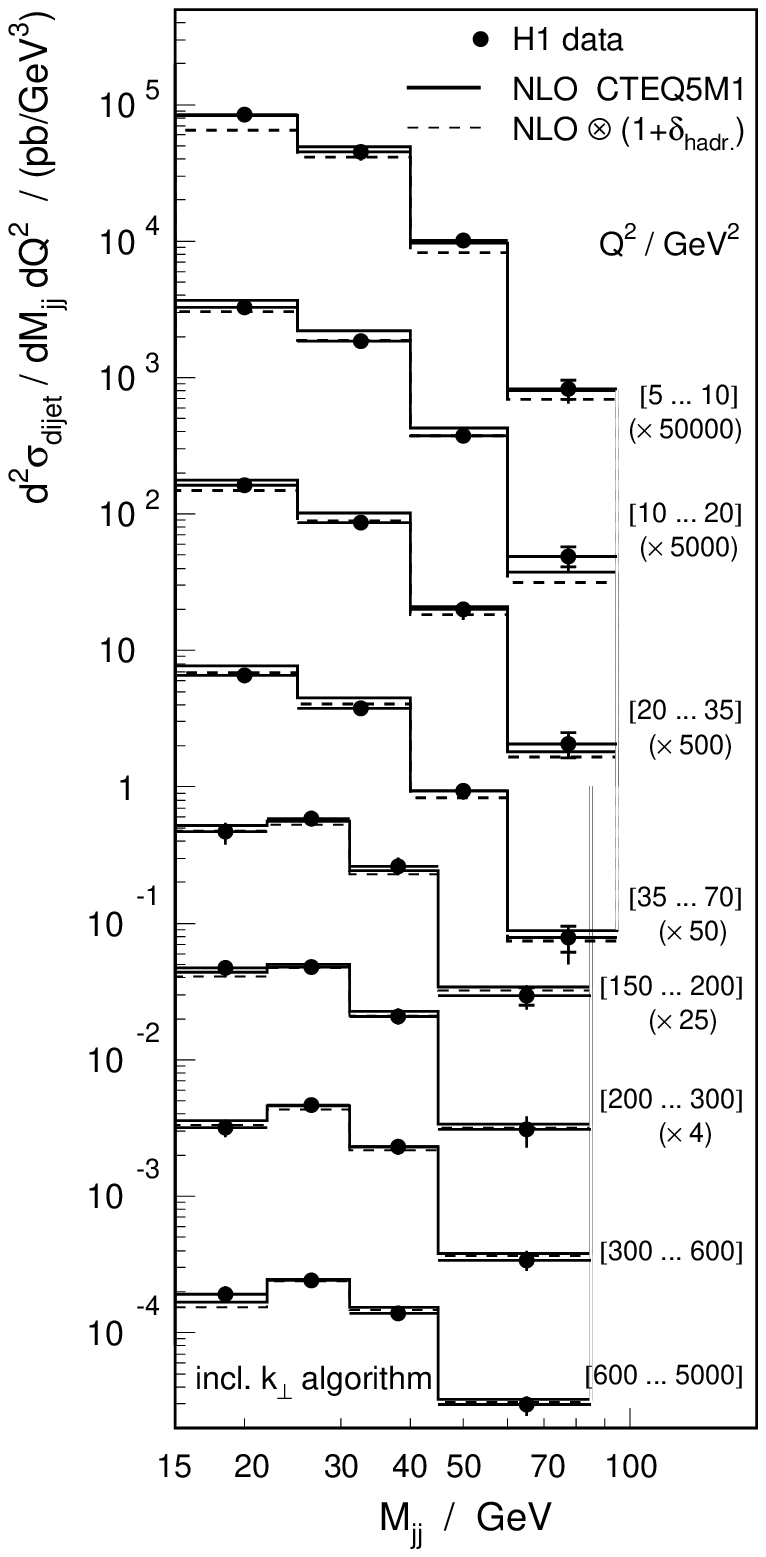,width=7.95cm}
 \epsfig{file=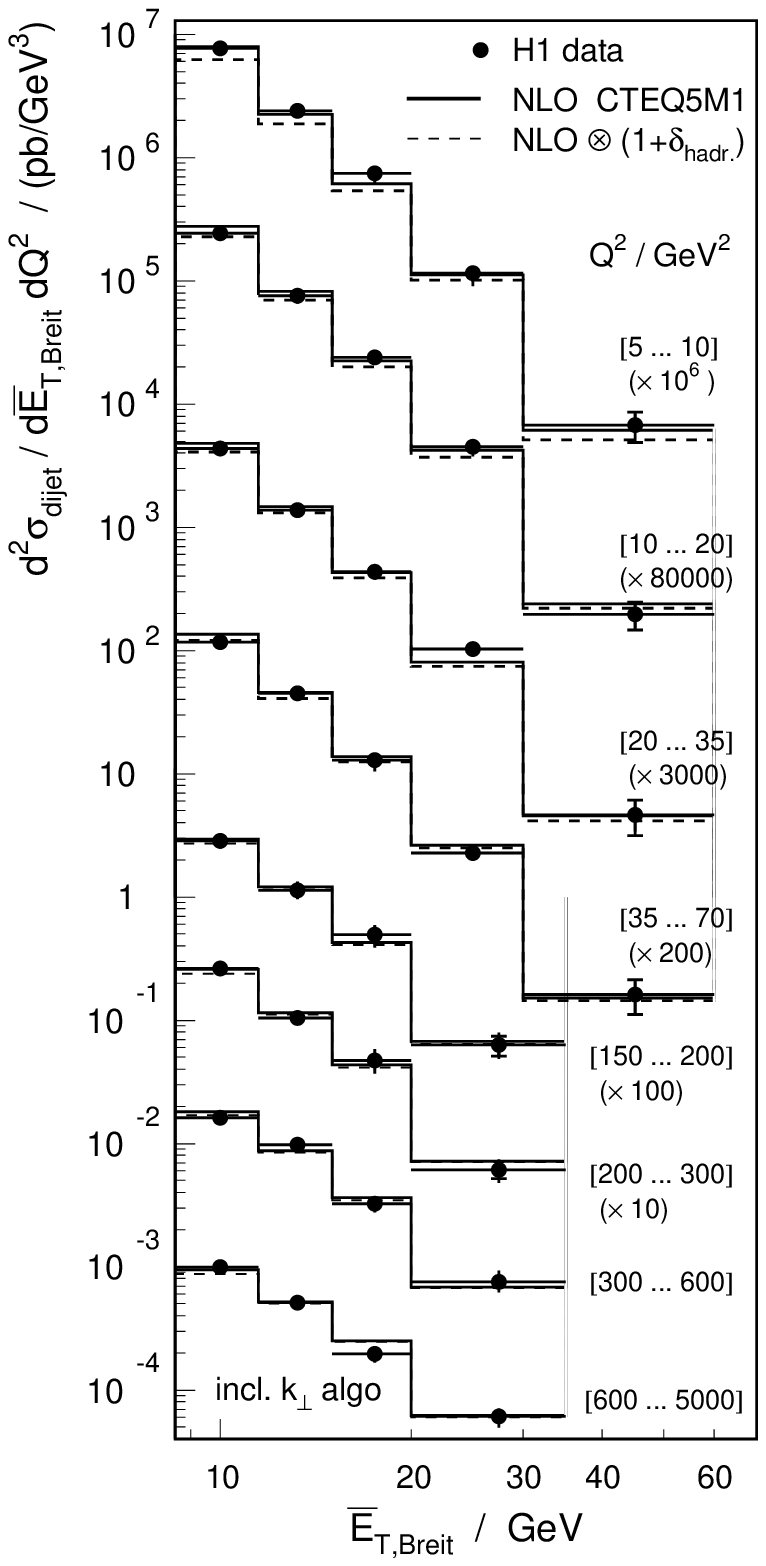,width=7.95cm}
\vskip-4mm
\caption{The dijet cross section for the inclusive $k_\perp$ 
algorithm as a function of $\mjj$ (left) and $\avet$ (right)
in different regions of $Q^2$.
The data are compared to the perturbative QCD prediction in NLO 
with (dashed line) and without (solid line) hadronization corrections
included.}
\label{fig:data_dij_etmjj} 
\end{figure}

\begin{figure}
\centering
 \epsfig{file=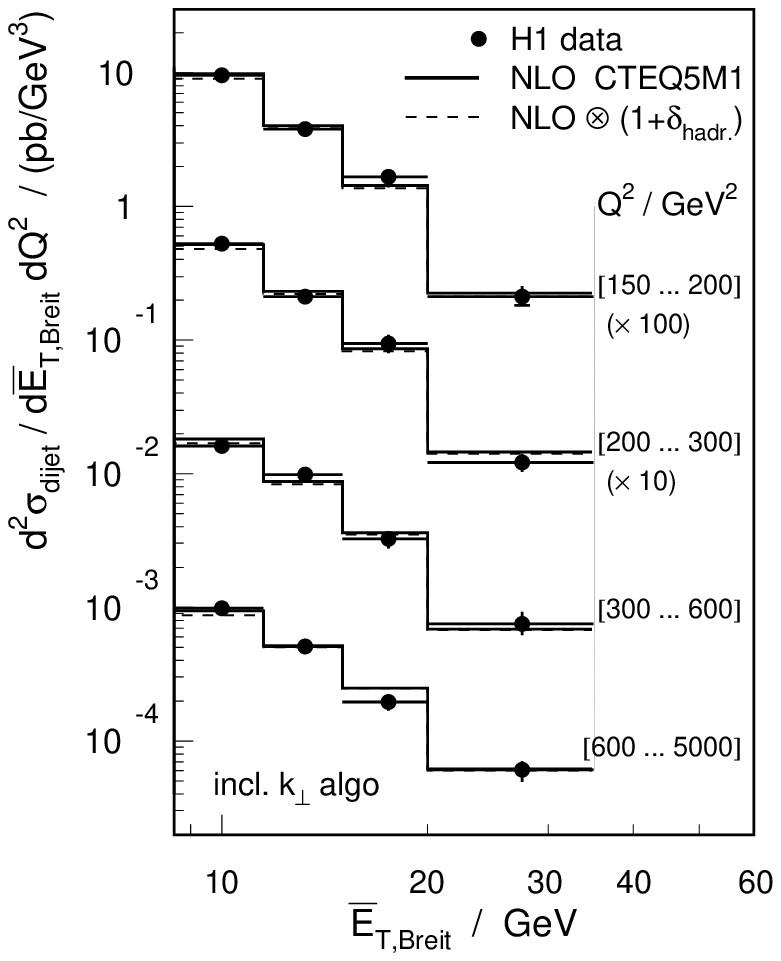,width=7.95cm}
 \epsfig{file=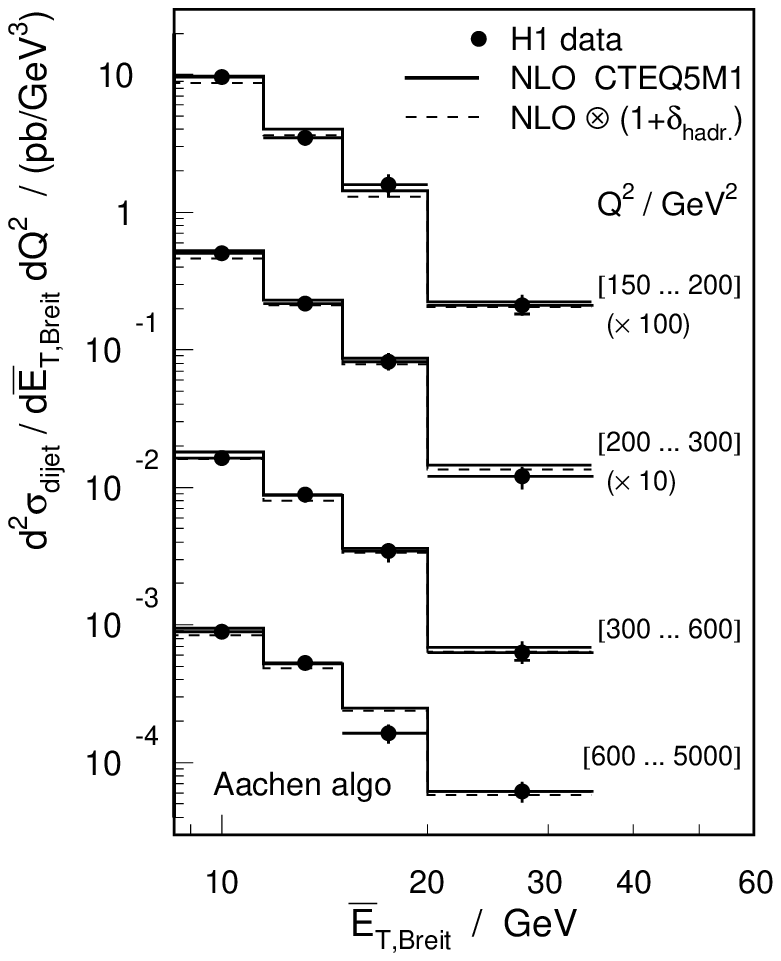,width=7.95cm}\vskip-1mm
 \epsfig{file=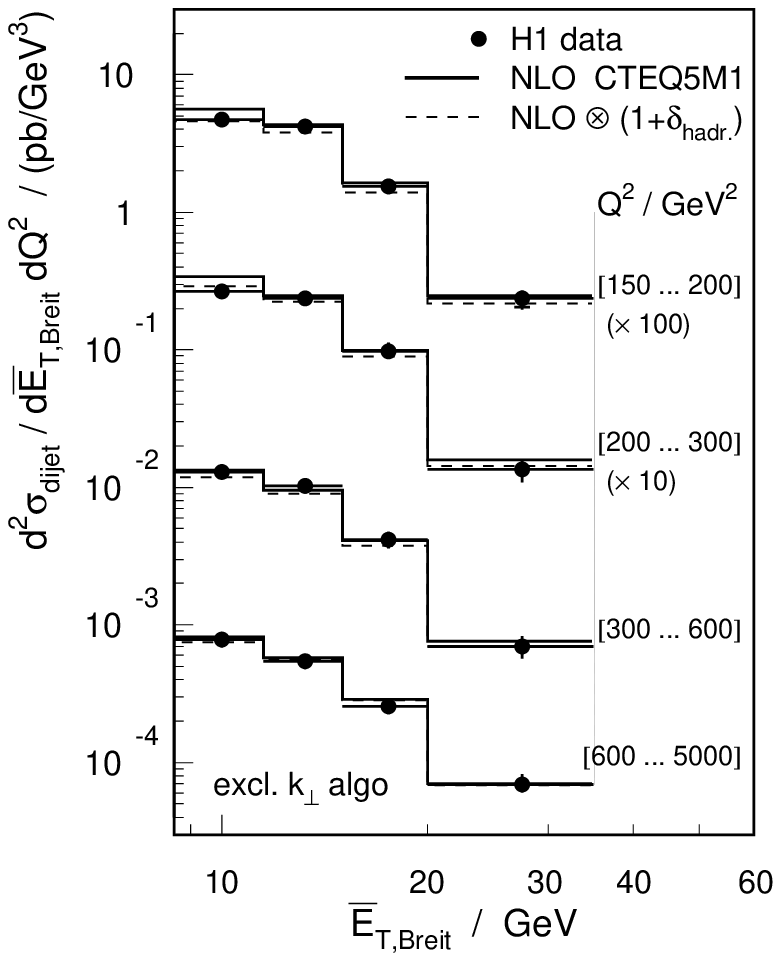,width=7.95cm}
 \epsfig{file=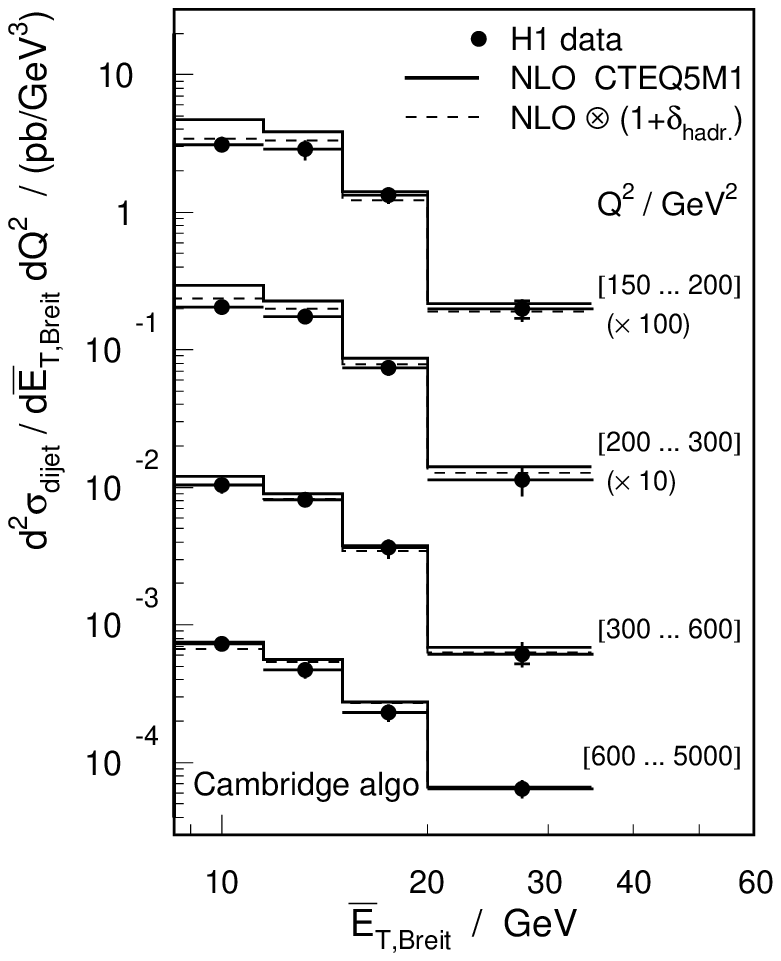,width=7.95cm}
\caption{The dijet cross section as a function of the average transverse
jet energy in the Breit frame in different regions of $Q^2$ for the 
inclusive $k_\perp$ algorithm (top left), the Aachen algorithm 
(top right), the exclusive $k_\perp$ algorithm (bottom left) and
the Cambridge algorithm (bottom right).
The data are compared to the perturbative QCD prediction in NLO 
with (dashed line) and without (solid line) hadronization corrections
included.}
\label{fig:data_dij_et} 
\end{figure}

The distributions of the invariant dijet mass $\mjj$
and the average transverse jet energy $\avet$
are shown in Fig.~\ref{fig:data_dij_etmjj} covering a range 
of $15 < \mjj < 95\GeV$ and $8.5 < \avet < 60\GeV$.
In both distributions we observe a harder spectrum
towards larger $Q^2$.
The NLO prediction, including hadronization corrections,
gives a good overall description, except at lowest $Q^2$
where it describes the shape, but not the magnitude, of the
cross section.
The increasing hardness of the $\avet$ distribution at higher 
$Q^2$ is also seen for the other jet algorithms in 
Fig.~\ref{fig:data_dij_et}.

\begin{figure}
\centering
 \epsfig{file=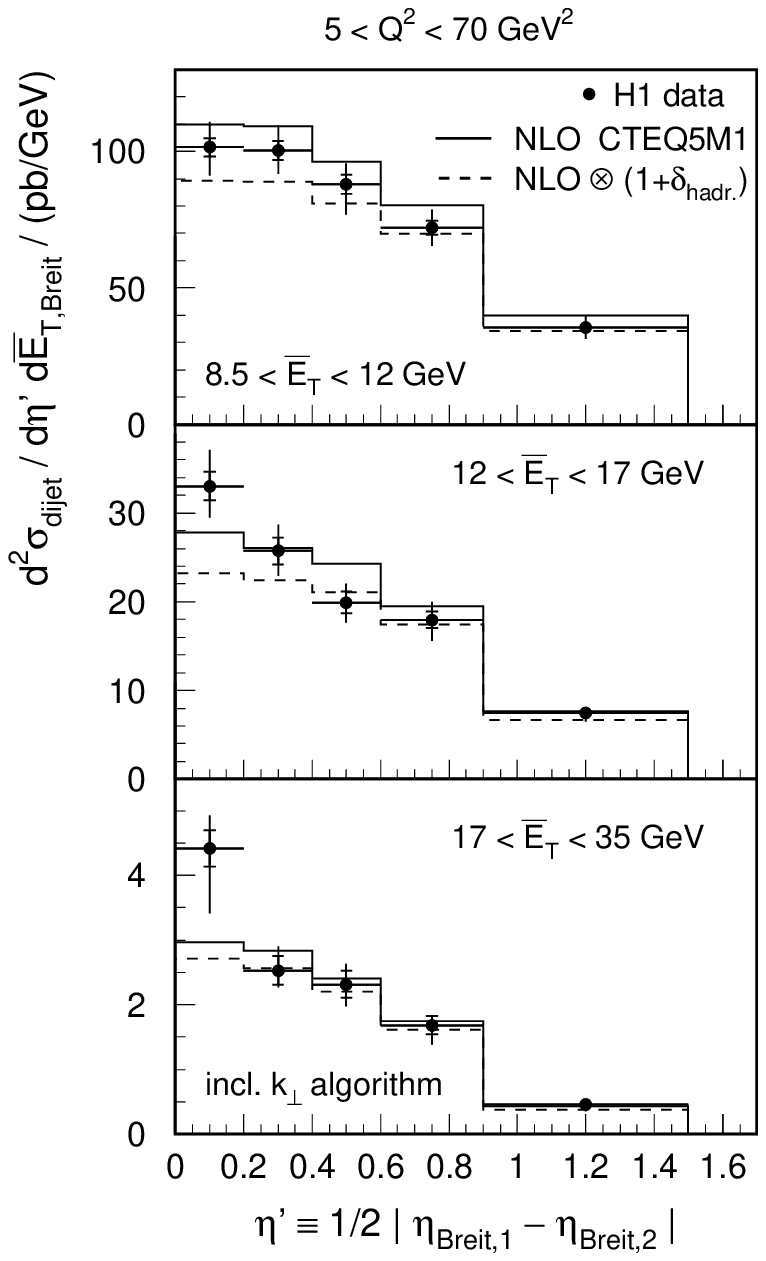,width=7.95cm}
 \epsfig{file=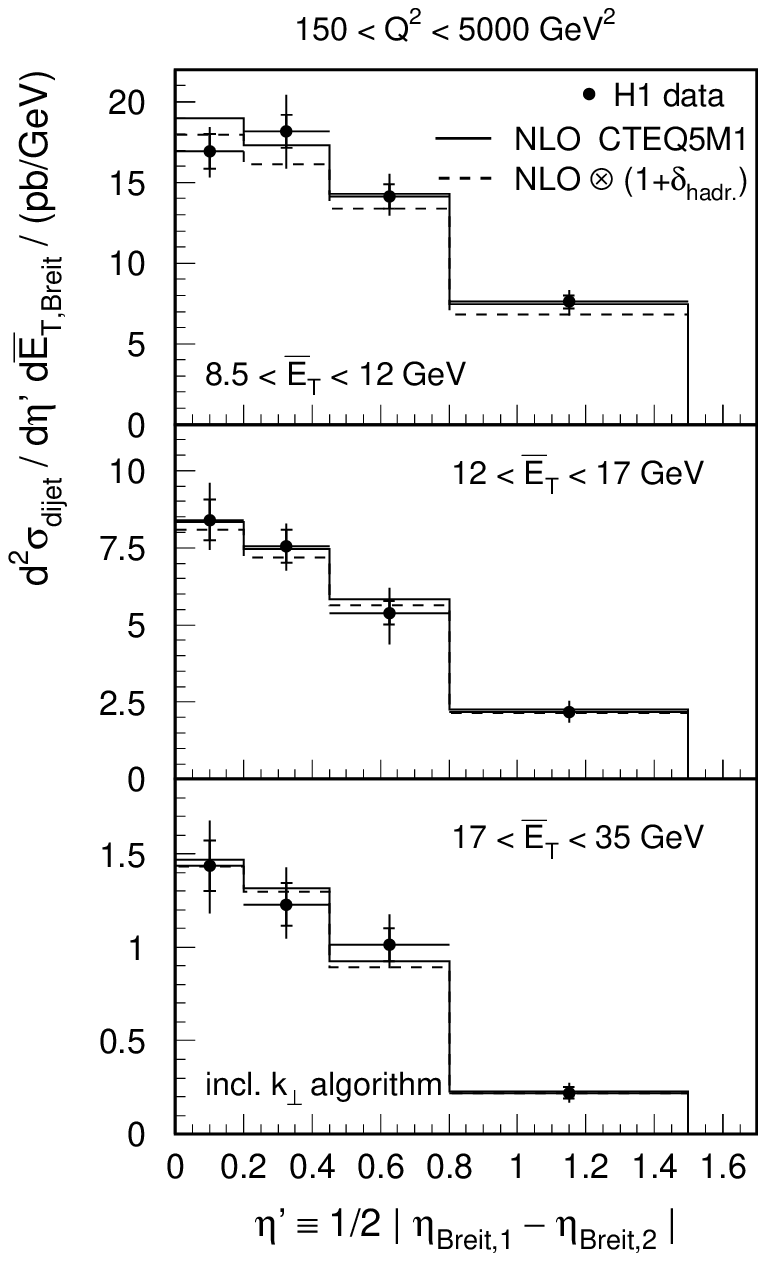,width=7.95cm}
\vskip-4mm
\caption{The dijet cross section for the inclusive $k_\perp$ 
algorithm as a function of the variable $\eta'$ in different
regions of $\avet$ at low $Q^2$ (left) and at high $Q^2$ (right).
The data are compared to the perturbative QCD prediction in NLO 
with (dashed line) and without (solid line) hadronization corrections
included.}
\label{fig:data_dij_etapq2} 
\end{figure}

The distribution of the pseudorapidity $\eta'$ 
(as defined in section~\ref{sec:psdef}) is shown in 
Fig.~\ref{fig:data_dij_etapq2} for different regions
of $\avet$ for the low and the high $Q^2$ data.
In both data sets the fraction of jets produced centrally in the 
dijet center-of-mass frame is observed to be larger
at higher $\avet$.

The partonic scaling variable $x_p$ is defined as the ratio
of the Bjorken scaling variable $\xbj$ and the reconstructed
parton momentum fraction $\xi$.
In the distribution shown in Fig.~\ref{fig:data_dij_xpxi} 
(left) a strong variation of the $x_p$ range is seen.
Towards lower $Q^2$ values $\xi$ differs by up to three orders of 
magnitude from $\xbj$.
At leading order the variable $\xi$ represents
the fraction of the proton momentum carried by the struck parton.
The dijet cross section in bins of $\xi$ is therefore directly
proportional to the size of the parton densities at the
parton momentum fraction $x=\xi$.
Fig.~\ref{fig:data_dij_xpxi} (right) shows the $\xi$ 
distribution in different regions of $Q^2$.
The dijet data are seen to be sensitive to partons with
momentum fractions $0.004 \lesssim \xi \lesssim 0.3$
which only increase slightly with increasing $Q^2$.
The $\xi$ distribution is of special importance in 
the QCD analysis for the determination of the gluon
density in the proton.
Therefore we display the contribution from 
gluon induced processes to this distribution which varies
strongly from $\simeq 80\%$ at low $Q^2$ to 
$\simeq 40\%$ at the highest $Q^2$.
Both, the $\xi$ and the $x_p$ distributions are well
described by the NLO calculation over the whole range 
of $Q^2$, independent of the fractional gluon contribution.

\begin{figure}
\vskip-1mm
\centering
 \epsfig{file=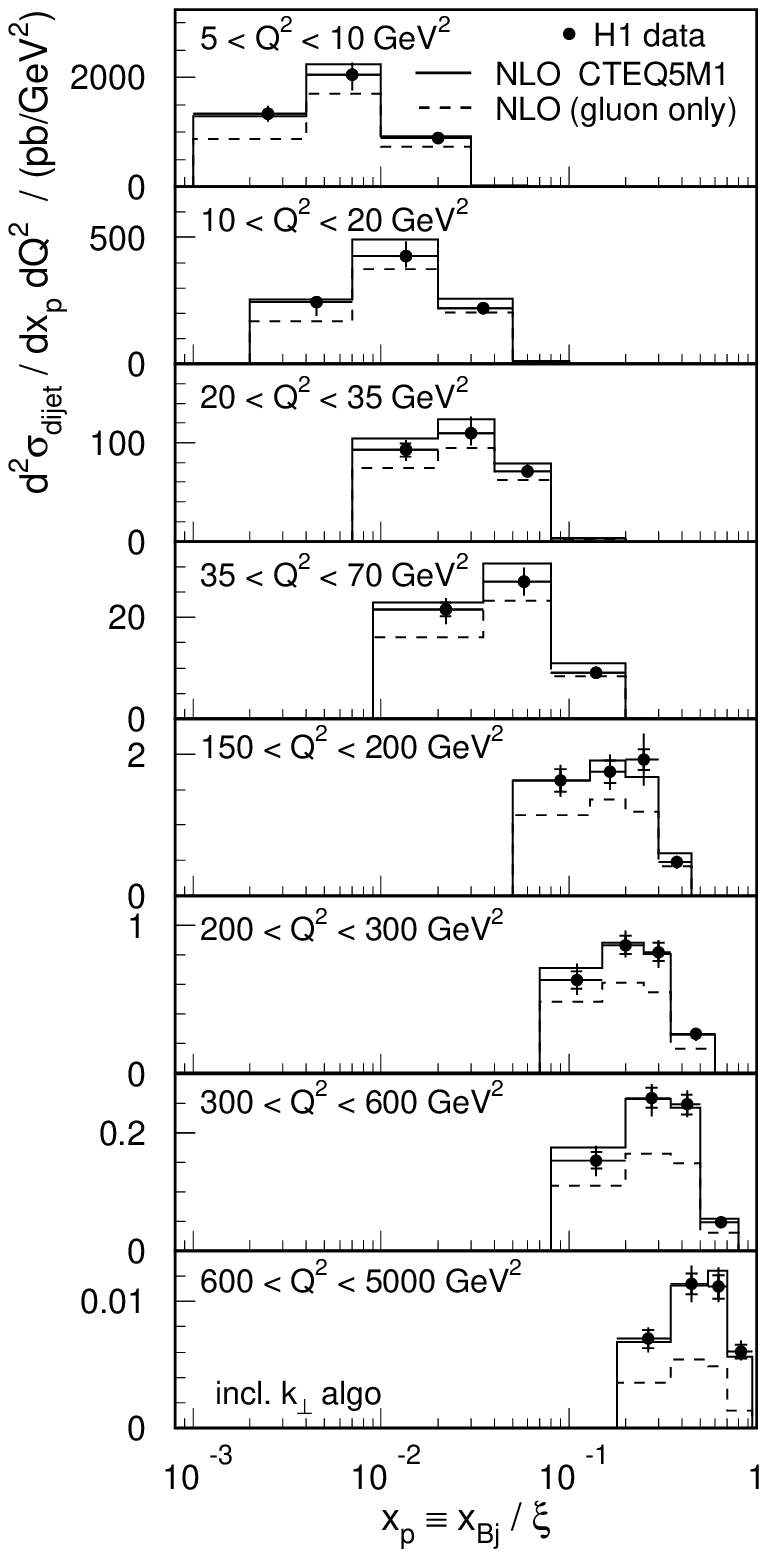,width=7.95cm}
 \epsfig{file=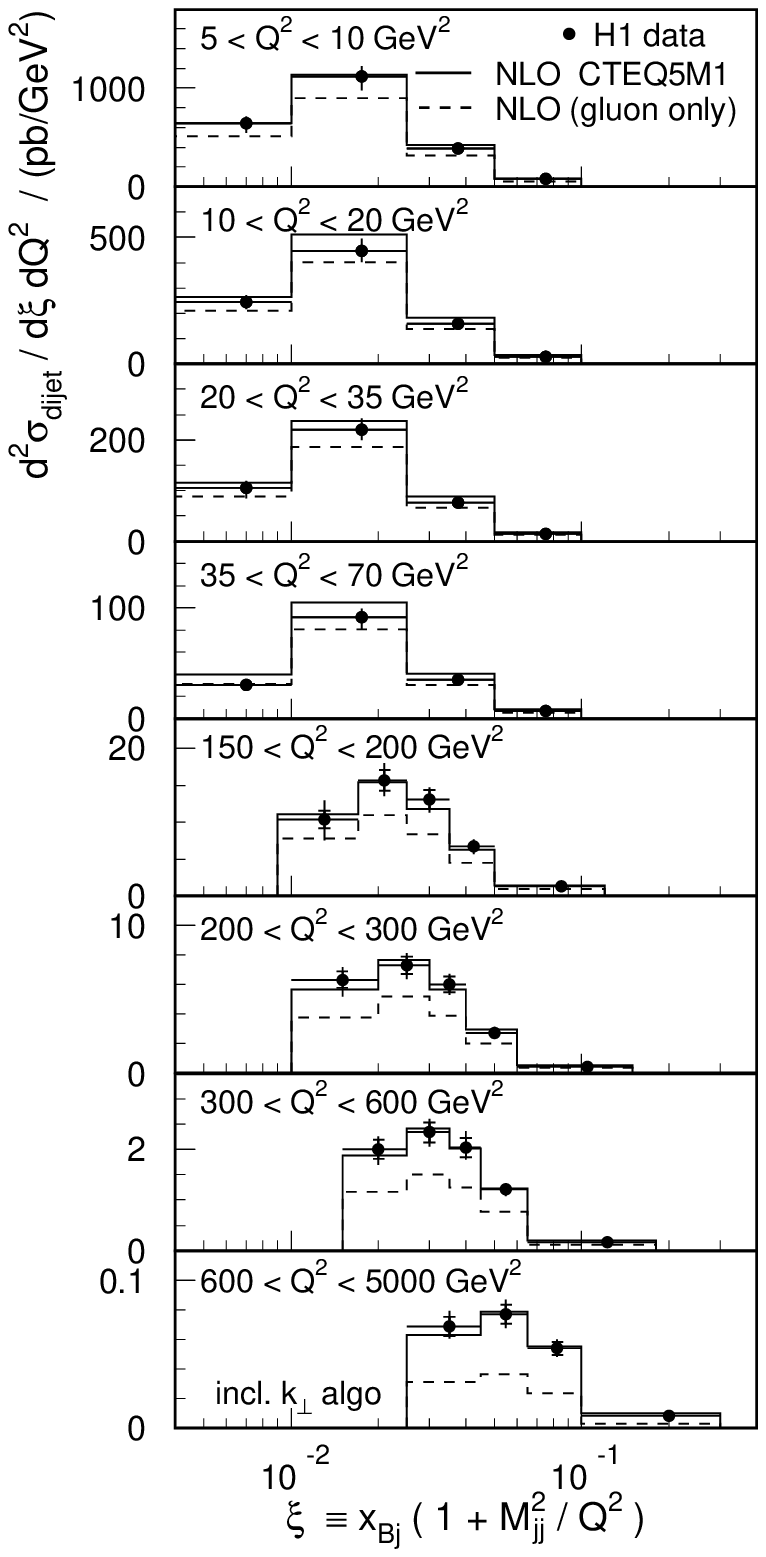,width=7.95cm}
\vskip-5mm
\caption{The dijet cross section for the inclusive $k_\perp$ 
algorithm as a function of the variables  $x_p$ (left)  and 
$\xi$ (right).
The perturbative QCD prediction in NLO (solid line) is 
compared to the measured dijet cross section. 
In addition the contribution from gluon induced processes is shown
(dashed line).}
\label{fig:data_dij_xpxi} 
\end{figure}

\begin{figure}
\centering
 \epsfig{file=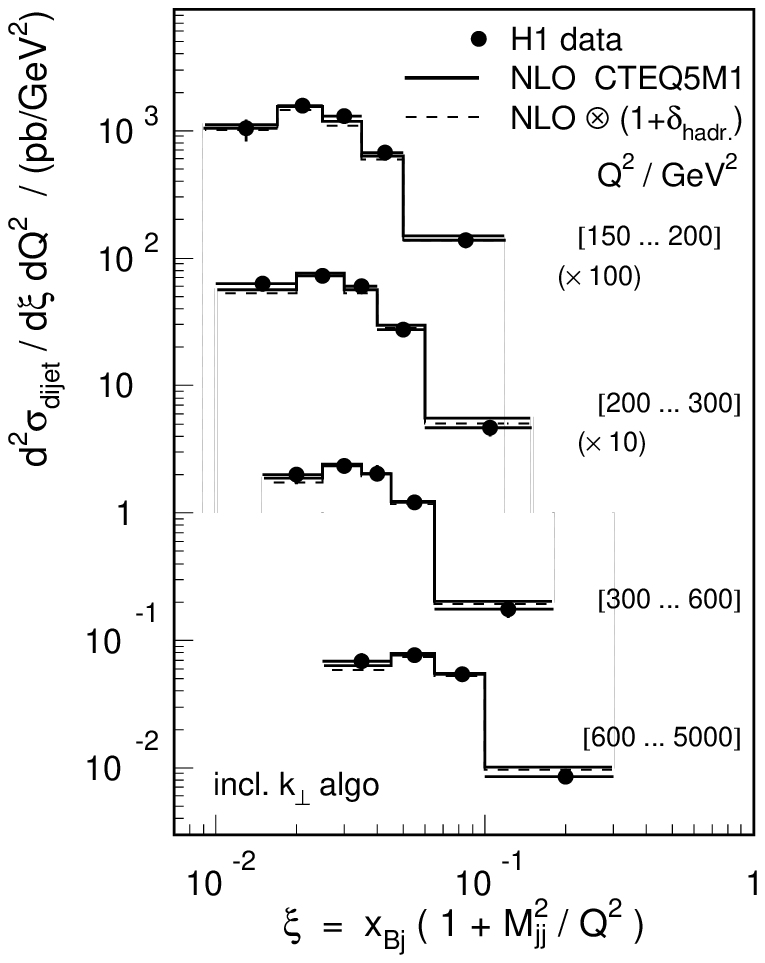,width=7.95cm}
 \epsfig{file=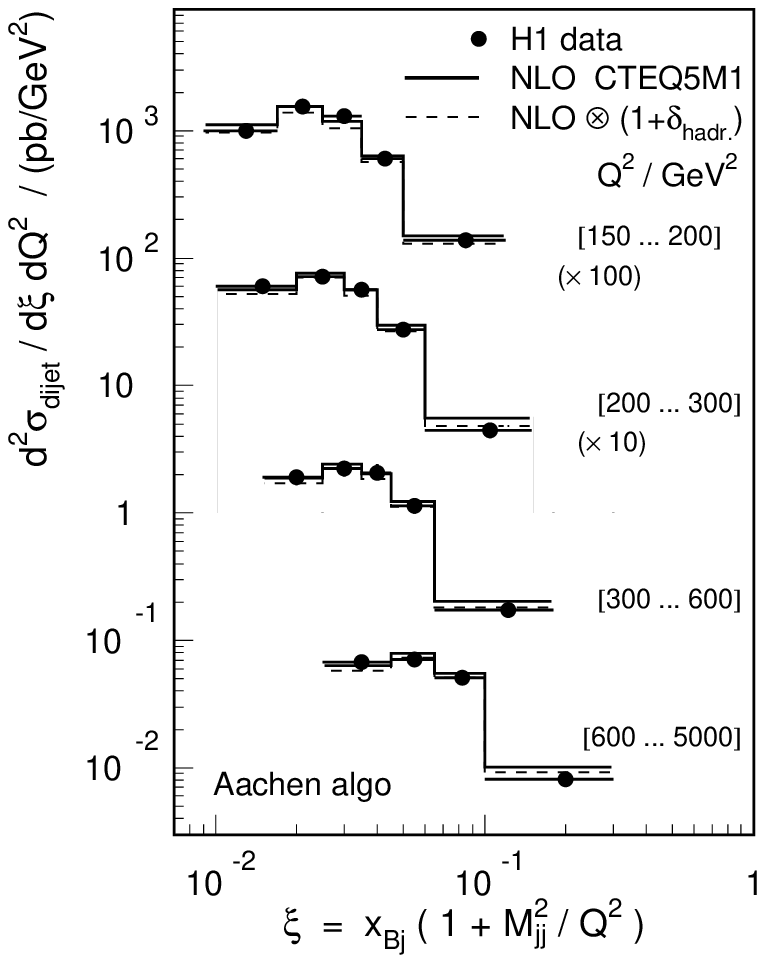,width=7.95cm}\vskip-1mm
 \epsfig{file=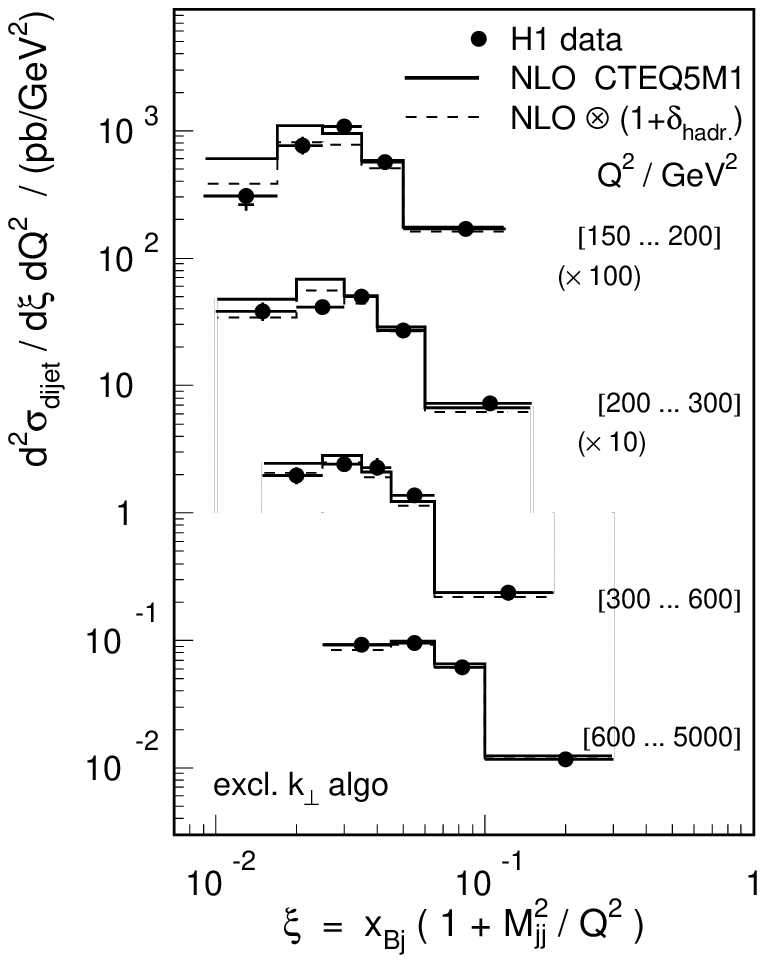,width=7.95cm}
 \epsfig{file=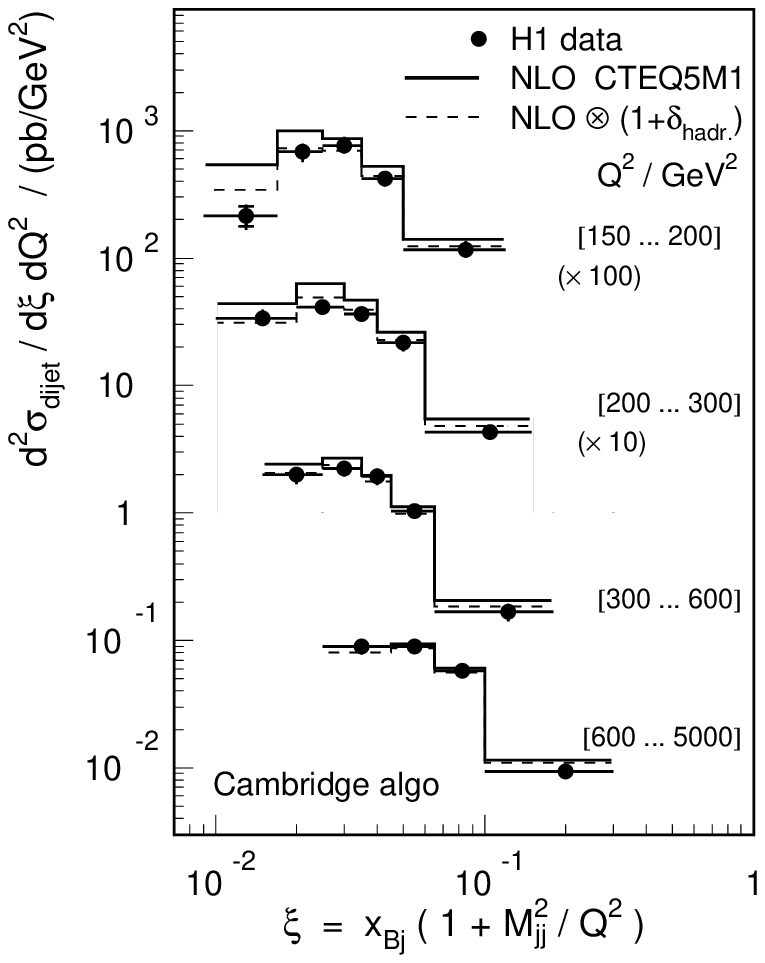,width=7.95cm}
\caption{The dijet cross section as a function of the 
reconstructed parton momentum fraction $\xi$.
The data are measured in different regions of $Q^2$ for the 
inclusive $k_\perp$ algorithm (top left), the Aachen algorithm 
(top right), the exclusive $k_\perp$ algorithm (bottom left) and
the Cambridge algorithm (bottom right).
The data are compared to the perturbative QCD prediction in NLO 
with (dashed line) and without (solid line) hadronization corrections
included.}
\label{fig:data_dij_xi} 
\end{figure}

\pagebreak[4]

The $\xi$ distribution is also presented for the other 
jet algorithms (Fig.~\ref{fig:data_dij_xi}).
While the distributions for the inclusive jet algorithms
(incl.\  $\kperp$ and Aachen) are already described
by the NLO calculation (without hadronization corrections),
large deviations are seen for the exclusive algorithms 
(excl.\  $\kperp$ and Cambridge), especially at small $\xi$
corresponding to small dijet masses.
However, in this region hadronization corrections are very
large for the exclusive algorithms.
Within the estimated size of these corrections
theory and data are consistent, except
in those regions where the corrections are especially large.

\begin{figure}
\centering
 \epsfig{file=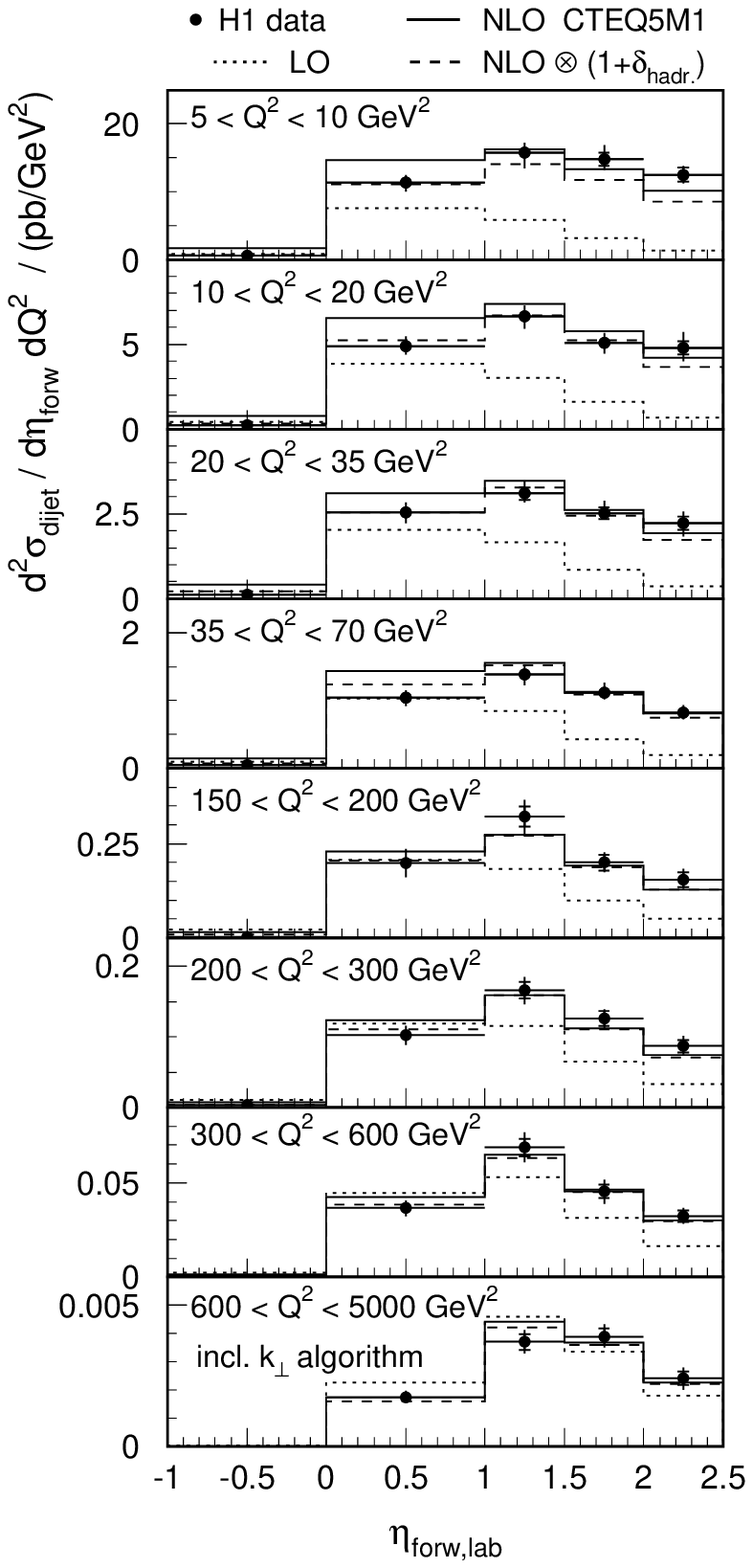,width=7.95cm}
\vskip-4mm
\caption{The dijet cross section for the inclusive $k_\perp$ 
algorithm as a function of the pseudorapidity of the 
forward jet in the laboratory frame.
The data are compared to the perturbative QCD prediction in NLO 
(solid line), in LO (dotted line)
and to a theoretical prediction where hadronization corrections
are added to the NLO calculation (dashed line).}
\label{fig:data_dij_etalab} 
\end{figure}

Fig.~\ref{fig:data_dij_etalab} finally shows the distribution
of the forward jet $\etaflab$ in the laboratory frame 
in different regions of $Q^2$.
While at larger $Q^2$ the distribution is 
seen to decrease towards the cut value at 
$\etaflab = 2.5$, it is flatter at low $Q^2$.
The theoretical calculation gives a reasonable
description of this angular distribution.
In addition also the LO prediction is included.
Although the NLO corrections become large in the forward region
(i.e.\  at large $\etaflab$) towards lower $Q^2$, the NLO calculation
does describe the data remarkably well.
Only at lowest $Q^2$ the NLO calculation clearly fails to describe 
the data, which is in agreement with the observations
made in an earlier analysis of forward jet 
production~\cite{h1fwdjet99}.

The perturbative NLO prediction
gives a good description of the data for those observables
for which NLO corrections and non-perturbative contributions
are small.
This agreement is seen in all regions of phase space,
independent of whether they are dominated by
the QCD-Compton or the boson-gluon fusion processes.
For observables with not too large hadronization corrections
the differences between the perturbative calculation
and the data can be explained by the predictions 
of phenomenological hadronization models.
In the kinematic region of $10< Q^2 < 70\GeV^2$
theory still gives a good description of the data
although NLO corrections become large.
The theoretical calculations only fail 
at $Q^2 < 10\GeV^2$ where NLO corrections
are largest (with $k$-factors above two),
such that contributions beyond NLO are expected to 
be sizable.

\clearpage
\section{QCD Analysis\label{sec:qcdana}}

The QCD predictions 
depend primarily on $\alps$ and on 
the gluon and the quark density functions of the proton.
In this section we present QCD analyses of the data
in which we determine these parameters of the theory.
We briefly discuss how different processes in DIS are directly
sensitive to the different parameters and introduce the physical 
and technical assumptions with which the QCD fits 
are performed.

\subsection{Strategy \label{sec:strategy}}

In perturbative QCD (pQCD) the cross section of any process 
in deep-inelastic lepton-proton scattering
can be written as a convolution of (process specific) 
perturbative coefficients $c_{a,n}$ with (universal) 
parton density functions $f_{a/p}$ of the proton
\begin{equation}
\sigma  \, = \,  
\sum_{a,n}  
\; \int_0^1 {\rm d}x \; \alpha_s^n(\mu_r) \;
 c_{a,n}\left(\frac{\xbj}{x}, \mu_r, \mu_f \right) \; f_{a/p}(x,\mu_f) \; .
\label{eq:directint}
\end{equation}
The sum runs over all contributing parton flavors $a$ 
(quarks and gluon) and 
all orders $n$ considered in the perturbative expansion.
The integration is carried out over all fractional parton 
momenta $x$.
The coefficients $c_{a,n}$ are predicted by pQCD.
They are currently known to next-to-leading order in the
strong coupling constant for the inclusive DIS cross section 
($n=0,1$) and for the dijet and the inclusive jet cross 
section ($n=1,2$)~\cite{zijlstra92}. 
In the regions of sufficiently large transverse jet energies
and not too large values of $Q^2$ ($Q^2 < 5\,000\GeV^2$)
the effects of $Z^\circ$ exchange and
of quark masses (for five quark flavors) can be neglected
as shown in~\cite{wobisch2000a} using the program
MEPJET~\cite{mepjet}.
In this approximation the perturbative coefficients of the quarks
for the inclusive DIS cross section and for the jet cross sections
fulfill the relations 
\begin{equation}
c_u = c_c \; = \; c_{\bar{u}} =  c_{\bar{c}}
\hskip6mm  \mbox{\small and} \hskip6mm 
c_d = c_s = c_b \; = \; c_{\bar{d}} =  c_{\bar{s}} =  c_{\bar{b}}  
\label{eq:coeff_relation}
\end{equation}
in each order of $\alps$.
Therefore only three coefficients are independent and the cross 
section in (\ref{eq:directint}) can be described by three independent
parton density functions\footnote{We do not explicitly display 
the dependence on the factorization scale.} 
$xG(x)$, $x\Delta(x)$ and $x\Sigma(x)$ with coefficients
$c_G$, $c_\Delta$ and $c_\Sigma$,
which have to be defined such that
\be
 c_g g(x) \, +\, \sum_a  c_a \, ( q_a(x) + \bar{q}_a(x) )
\; = \; 
c_G \, G(x) \: +\: c_\Sigma \,\Sigma(x)\: +\: c_\Delta\,\Delta(x) \: ,
\end{equation}
where the sums run over all quark flavors $a$.
These three parton density functions are chosen to be
\begin{eqnarray}
\mbox{Gluon:} \hskip7mm
  x \, G(x) & \equiv & x \, g(x) \, , \nonumber \\
\mbox{Sigma:} \hskip7mm
  x \, \Sigma(x)  & \equiv & x \, \sum_a \, ( q_a(x) + \bar{q}_a(x) ) \, ,
\nonumber \\
\mbox{Delta:} \hskip7mm
  x \, \Delta(x) & \equiv & x \, 
\sum_a  e_a^2 \; ( q_a(x) + \bar{q}_a(x) ) \, ,
\label{eq:def_pdf}
\end{eqnarray}
where  $e_a$ denotes the 
electric charge of the corresponding quark.
The three corresponding coefficients are given by linear 
combinations of the single flavor coefficients
\be
c_G  =  c_{\rm gluon}               \hskip11mm
c_\Sigma  =  \frac{1}{3} \; (4\, c_d - c_u) \hskip11mm
c_\Delta  =  3 \; (c_u - c_d)   \; .
\label{eq:def_coeff}
\end{equation}

At orders ${\cal O}(\alpha_s^0)$ and ${\cal O}(\alpha_s^1)$
the contributions from different quark flavors are
proportional to their electric charge squared
(i.e.\  $c_u = 4\, c_d$).
Therefore
the coefficient $c_\Sigma$ in (\ref{eq:def_coeff}) vanishes 
and the only quark contributions to the cross sections come from 
$x\Delta(x)$.
The gluon gives contributions 
at order ${\cal O}(\alps^1)$ and higher.
$x\Sigma(x)$ starts to contribute at order 
${\cal O}(\alpha_s^2)$ and does therefore not enter the 
inclusive DIS cross section to next-to-leading order.
Table~\ref{tab:pdforders} gives an overview of the orders 
in which the parton densities contribute to the different 
processes (to NLO).
\begin{table}[t]
\centering
\begin{tabular}{|l|c|c|}
\hline 
      & \sf LO & \sf NLO \\ 
\hline
$\sigma_{\rm incl.\ DIS}$ & $x\Delta(x)$ & $x\Delta(x), \; xG(x)$ \\
$\sigma_{\rm jets}$  & \,\,  $xG(x), \; x\Delta(x)$ \,\, & 
\,\, $xG(x), \; x\Delta(x), \; x\Sigma(x) $ \,\, \\
\hline
\end{tabular}
\caption{Overview of the parton density functions contributing 
to different cross sections in LO and NLO. \label{tab:pdforders}}
\end{table}
$x\Sigma(x)$ enters only the jet cross sections via the 
NLO corrections.
At large $Q^2$ the contributions from $x\Sigma(x)$ are, however, 
small (4.5\% for the dijet cross section at $150 < Q^2 < 200\GeV^2$,
decreasing to 2\% at $600 < Q^2 < 5000\GeV^2$).
In the following the parameterization CTEQ5M1 is used to 
determine this contribution which is not regarded 
as a degree of freedom in the analysis.
This is, however, only a weak assumption which will 
(due to the smallness of the contribution)
not bias the result.

With this approximation
the inclusive DIS cross section and the jet cross section
now depend on three quantities which will be determined
in this analysis:
$\alps$, the gluon density $xG(x)$ and the 
quark density $x\Delta(x)$.
To demonstrate the basic sensitivity the leading
order cross sections are written in the symbolic form
\bea
\mbox{inclusive DIS cross section:} \hskip8mm 
\sigma_{\rm incl.\, DIS} & \propto &  \Delta    \nonumber \\
\mbox {jet cross sections in DIS:} \hskip20.5mm 
\sigma_{\rm jet} & \propto & 
\alps \cdot ( c_G  \, G \; + \; c_\Delta \, \Delta  )      \, .
\label{eq:fiteq}
\eea
These relations make clear
that in DIS a direct determination of either $\alps$ or the 
gluon density can never be performed without considering
the correlation with the other quantity.
Three strategies are used in the QCD fits which differ 
by the amount of external information included in the analysis.

\begin{enumerate}

\item Determination of $\alps$ from jet cross sections:
using the jet cross sections measured one can determine
$\alps$ assuming external knowledge on the parton distributions
as provided by global data analyses.

\item Consistent determination of the gluon density $xG(x)$ and
the quark density $x\Delta(x)$:
including data on the inclusive DIS cross section,
which are directly sensitive to the quarks only, and
assuming the world average value of $\alpsmz$
the information provided by the jet data can be used
for a direct determination of the gluon density together 
with the quark densities via a
simultaneous fit.

\item Simultaneous determination of $\alps$, 
the gluon density $xG(x)$ and the quark density $x\Delta(x)$:
if the jet cross sections are measured in different phase 
space regions ($\sigma_{\rm jet}, \, \sigma_{\rm jet}'$)
with different sensitivity to the quark and the gluon 
contributions (i.e.\  where $c_G'/c_\Delta' \ne c_G/c_\Delta$)
a simultaneous direct determination of all free parameters is
possible when again additional inclusive DIS data are included.

\end{enumerate}

\subsection{Fitting Technique\label{sec:fittec}}

A determination of theoretical parameters can only be performed
in phase space regions where theoretical predictions are reliable.
Although the perturbative NLO calculation gives a good description
of the jet data down to $Q^2 = 10\GeV^2$, the QCD analysis is 
restricted to the region where NLO corrections are small
(with $k$-factors below $1.4$), i.e.\
to the region of high $Q^2$ ($150 < Q^2 < 5\,000\GeV^2$).
For the main analysis the jet cross sections measured
for the inclusive $\kperp$ algorithm are used for which hadronization 
corrections are smallest.
Jet cross sections from other jet algorithms are used to test the 
stability of the results.
The uncertainties of the jet data and their correlations
are treated as described in section~\ref{sec:expunc}.

In the second and in the third step of the analysis 
data on the (reduced) inclusive DIS cross section 
are included to exploit their
sensitivity to the quark densities in the proton.
A subsample is taken of the recently published 
measurement~\cite{h1highq2} in the range 
$150 \le Q^2 \le 1\,000\GeV^2$.
Since the present analysis uses the same experimental techniques
the effects of the point to point correlated experimental 
uncertainties can be fully taken into account.

The fit of the theoretical parameters is performed in a $\chi^2$ 
minimization using the program MINUIT~\cite{minuit}.
The definition of $\chi^2$~\cite{zomer95} fully takes into 
account all correlations of experimental and theoretical
uncertainties.
This $\chi^2$ definition has also been used in 
recent global data analyses~\cite{botje99,zomer99} and
in a previous H1 publication~\cite{h1compositeness99}.
The quoted uncertainties of the fit parameters are defined by 
the change of the parameter for which the $\chi^2$ of the fit 
is increased by one.

In the fitting procedure the perturbative QCD predictions
in NLO for the inclusive DIS cross section are directly compared
to the data, 
while the NLO predictions for the jet cross sections are
corrected for hadronization effects
before they are compared to the jet data:
\begin{eqnarray}
\sigma^{\rm H1}_{\rm incl. \,DIS}  & \longleftrightarrow &
          \sigma^{\rm NLO}_{\rm incl. \,DIS}  \nonumber \\
\sigma^{\rm H1}_{\rm jet}  & \longleftrightarrow &
          \sigma^{\rm NLO}_{\rm jet} \cdot
( 1+ \delta_{\rm  hadr.} )
 \hskip10mm \mbox{with}  \hskip5mm 
 \delta_{\rm  hadr.} = 
\frac{\sigma_{\rm jet}^{\rm hadron} - \sigma_{\rm jet}^{\rm parton}}%
{\sigma_{\rm jet}^{\rm parton}}
 \; .  \nonumber 
\end{eqnarray}
The hadronization corrections are determined as described
in section~\ref{sec:theocalc} using the average value from
the model predictions by HERWIG, LEPTO and ARIADNE.
The uncertainty from the model and the parameter dependence
of these predictions is always 
below 3\%~\cite{wobisch2000a,wobisch99b}.
The uncertainty in the matching of the parton level
(parton cascade and NLO calculation) is taken 
into account by increasing the quoted uncertainty in those regions
where the hadronization corrections are large.
In detail, the uncertainty of the hadronization correction for 
each bin of the jet cross sections is taken to be half 
the size of the correction, but at least 3\%.
This uncertainty is assumed to be correlated between the
theoretical predictions for all data points.

The renormalization scale $\mu_r$ in the NLO calculation
is identified with the process specific hard scales in both 
processes.
The inclusive DIS cross section is evaluated at $\mu_r = \sqrt{Q^2}$ 
and the jet cross sections are evaluated at $\mu_r = E_T$
(inclusive jet cross section) and  $\mu_r = \avet$ 
(dijet cross section).
For the jet cross section an example is given of how the results
change for an alternative choice, $\mu_r = \sqrt{Q^2}$.
The strong coupling constant $\alps(\mu_r)$ is parameterized 
in terms of its value at the scale $\mu_r = M_Z$ using the 
numerical solution of the renormalization group equation 
in 4-loop accuracy\footnote{It has been checked
that in the range of scales considered in this analysis,
$7\GeV < \mu_r < M_Z$, the differences
between the 2-, 3- and 4-loop solutions are always 
below 3 per mille.}~\cite{4loop,4loopmatch}.

In principle the arguments invoked in the choice of the 
renormalization scale $\mu_r$ also apply to the 
factorization scale $\mu_f$ for the inclusive DIS cross section 
and for the jet cross section.
However, a different choice is made for the following reasons.
The different parton flavors have been combined into
three independent parton density functions $xG(x,\mu_f)$, $x\Delta(x,\mu_f)$ 
and $x\Sigma(x,\mu_f)$ (section~\ref{sec:strategy}).
These three parton densities are, however, only independent as 
long as no evolution between different scales $\mu_f$ is performed.
The evolution of the gluon density is coupled to the evolution of 
$x\Sigma(x,\mu_f)$.
Furthermore, since $x\Delta(x,\mu_f)$ is not an eigenstate of the DGLAP 
evolution operators, the evolution requires its decomposition into a
non-singlet and a singlet (i.e.\  $x\Sigma(x,\mu_f)$).
This introduces an additional dependence between the 
quark densities.
To avoid mixing between the different parton densities
the parton distributions are not evolved to different scales.
Instead the perturbative calculations are carried out at a fixed 
value of the factorization scale $\mu_f = \mu_0$.
The jet cross sections are sensitive to the parton distributions
in the $x$-range  $0.008 \lesssim x \lesssim 0.3$ 
(see Fig.~\ref{fig:data_dij_xpxi}).
In this $x$-range the factorization scale dependence of the parton 
density functions is not large.
In a next-to-leading order calculation
the remaining $\mu_f$ dependence given by the DGLAP 
evolution equations is largely compensated by
a corresponding term 
in the perturbative coefficients. 
The perturbative cross sections therefore depend only weakly
on the choice of the factorization scale.
The difference between using a fixed factorization scale $\mu_0$
and performing the full DGLAP evolution at a scale $\mu_f$
is of higher order in $\alps$ 
than those considered.
If the scale $\mu_f$ is close to the fixed scale 
these higher order terms which are proportional to 
$\ln(\mu_f/\mu_0)$ are small.
Therefore a fixed value of the factorization scale
of the order of the average transverse jet energies
in the dijet and the inclusive jet cross section
$\mu_f = \mu_0 = \sqrt{200}\GeV \simeq \langle E_T \rangle$
is used.
The subsample of the (reduced) inclusive DIS cross section
$150 \le Q^2 \le 1\,000\GeV^2$ has been chosen such that the 
four-momentum transfer is also of the same order 
of magnitude $\sqrt{Q^2} \simeq \mu_0 = \sqrt{200}\GeV$.

Both the renormalization and the factorization scale dependences
of the cross sections each are considered as correlated
theoretical uncertainties.
Both scales are (separately) varied by a factor $x_\mu$ 
around their nominal values $\mu_0$ in the range 
$x_\mu= \{ \frac{1}{2},2\}$ and the ratios
$\frac{\sigma_{\rm NLO}(x_\mu \cdot \mu_0) }{\sigma_{\rm NLO}(\mu_0)}$
are taken as the corresponding uncertainties.
Together with the uncertainty from the hadronization
corrections they constitute the quoted theoretical 
uncertainty of the fit results.

During the $\chi^2$ minimization procedure in the fit the 
NLO calculations of the jet cross sections have to be performed 
iteratively for different values of $\alpsmz$ and for different 
parton density functions (the number of calculations used to obtain 
the present results and to study their stability is in the order 
of one million).  
Since standard computations of NLO jet cross sections are time 
consuming the method~\cite{wobisch2000a} is used of 
pre-convoluting the perturbative coefficients with suitably 
defined functions which can then be folded with 
the parton densities and $\alps$ for a
fast computation of the NLO cross section.

\subsection{\boldmath Determination of $\alps$ \label{sec:fitalps}}

As a first step the QCD predictions are fitted to the jet 
cross sections using parameterizations for the parton 
distributions from global fits.
The single free parameter which is determined in the fits 
is the value of the strong coupling constant.
All $\alps$ fit results presented hereafter consider
all experimental and theoretical uncertainties.
The contribution of the uncertainties of the parton distributions 
to the uncertainty of $\alps$ is  discussed separately.

The value of $\alpsmz$ is obtained from a fit to the
inclusive jet cross section
measured double-differentially with the inclusive $k_\perp$ algorithm.
For the result, we use the parton distributions from the
CTEQ5M1 parameterization~\cite{cteq5} and check the effects of 
other choices.
The renormalization scale is chosen to be $\mu_r = E_T$ 
and the factorization scale
is set to the fixed value of $\mu_f = \sqrt{200}\GeV$ 
(the average $E_T$ of the jet sample).
The effect on $\alpsmz$ of using a different choice for $\mu_r$
is studied.

The studies of the stability of the results include fits
to the inclusive jet cross section measured with the Aachen 
jet algorithm, fits to the double-differential dijet cross section
\linebreak
${\rm d}^2 \sigma_{\rm dijet} / {\rm d}\overline{E}_T {\rm d}Q^2$ using four 
different jet algorithms, and fits to other double-differential
dijet distributions.

\subsubsection*{Fits to Single Data Points}

\begin{figure}[t!]
\centering
 \epsfig{file=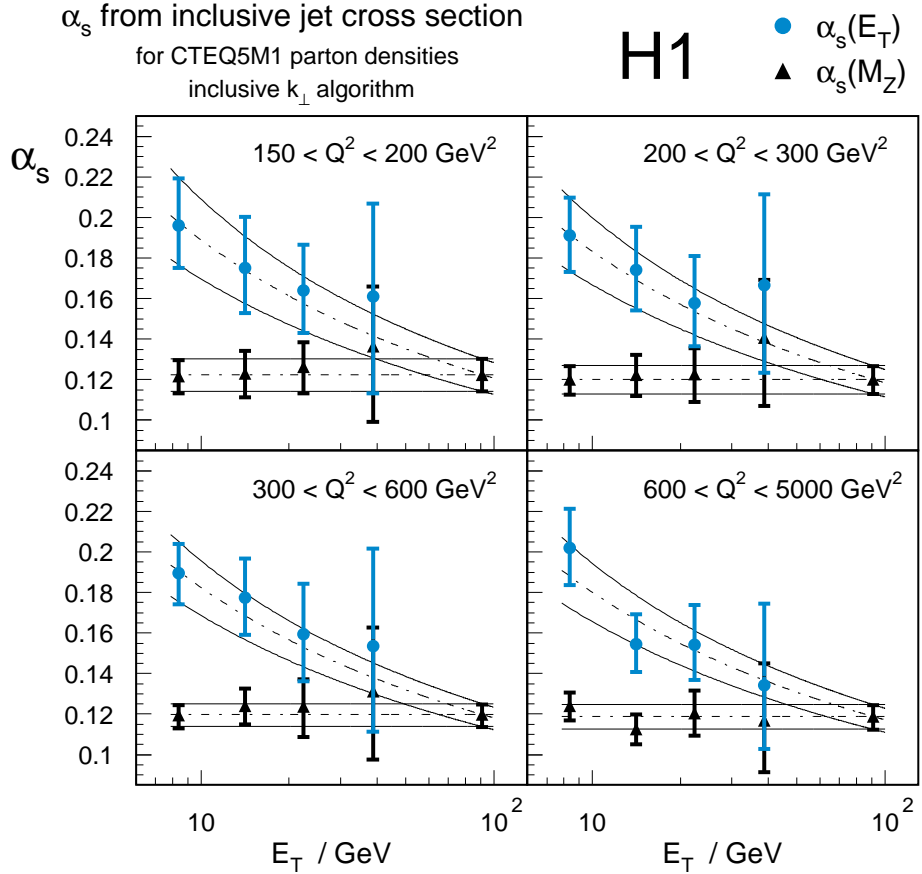,width=12.5cm}
\vskip-2mm
\caption{Determination of $\alpha_s$ from the inclusive jet cross section
using the inclusive $k_\perp$ algorithm at
a renormalization scale $\mu_r = E_T$ .
Displayed are the results of the fits to the single data 
points in each $Q^2$ region at each $E_T$ value (circles)
including experimental and theoretical uncertainties.
The single values are extrapolated to the $Z^0$ mass (triangles).
A combined fit yields a result for $\alpha_s(M_Z)$ 
(rightmost triangle) for each $Q^2$ region.
The lower curves represent the combined fit results and their 
uncertainties and the upper curves indicate the prediction of the 
renormalization group equation for their evolution.}
\label{fig:q2results}
\end{figure}

Before carrying out combined fits to groups of data points
the consistency of the data is tested by performing QCD fits 
separately to all sixteen single data points of the double-differential 
inclusive jet cross section.

The fit results are displayed in Fig.~\ref{fig:q2results}
for the four regions of $Q^2$.
In each fit a result for $\alpset$ is extracted which is
presented  at the average $E_T$ of the corresponding data point.
The individual results are subsequently evolved to $\alpsmz$.
Combined fits to all four data points in the same
$Q^2$ regions are performed, leading to a combined result
of $\alpha_s(M_Z)$ for each $Q^2$ region.
The lower curves in the plots represent the combined fit results 
and their uncertainties and the three upper curves indicate the 
evolution of the combined result and its uncertainty according to the 
renormalization group equation.
The single $\alpset$ values are consistent with the predicted 
scale dependence of $\alps$ and all combined $\alpsmz$ results 
are compatible with each other.
The results obtained in the different $Q^2$ regions are 
(for $\mu_r = E_T$)
\bea
150 < Q^2 < 200\GeV^2 \, :  \;& \alpsmz = &  0.1225 
    \, ^{+0.0052}_{-0.0054} \; ({\rm exp.})  
    \, ^{+0.0060}_{-0.0062} \; ({\rm th.})  \, , \nonumber \\
200 < Q^2 < 300\GeV^2 \, :  \;& \alpsmz = & 0.1202 
    \, ^{+0.0044}_{-0.0044} \; ({\rm exp.}) 
    \, ^{+0.0052}_{-0.0056} \; ({\rm th.}) \, , \nonumber \\
300 < Q^2 < 600\GeV^2 \, :  \;& \alpsmz = &  0.1198
    \, ^{+0.0037}_{-0.0038} \; ({\rm exp.}) 
    \, ^{+0.0040}_{-0.0046} \; ({\rm th.}) \, , \nonumber \\
600 < Q^2 < 5000\GeV^2 \, :  \;& \alpsmz = &  0.1188
    \, ^{+0.0048}_{-0.0048} \; ({\rm exp.}) 
    \, ^{+0.0035}_{-0.0042} \; ({\rm th.}) \, . 
\eea
While the experimental uncertainties are of similar size
for all $\alpsmz$ values, the theoretical uncertainties
shrink slightly towards larger $Q^2$.
This is a consequence of the reduced renormalization scale 
dependence of the jet cross section at higher $Q^2$.

\subsubsection*{\boldmath Combined Fit -- Central $\alpsmz$ 
Result \label{sec:as_et}}

\begin{figure}
\centering
 \epsfig{file=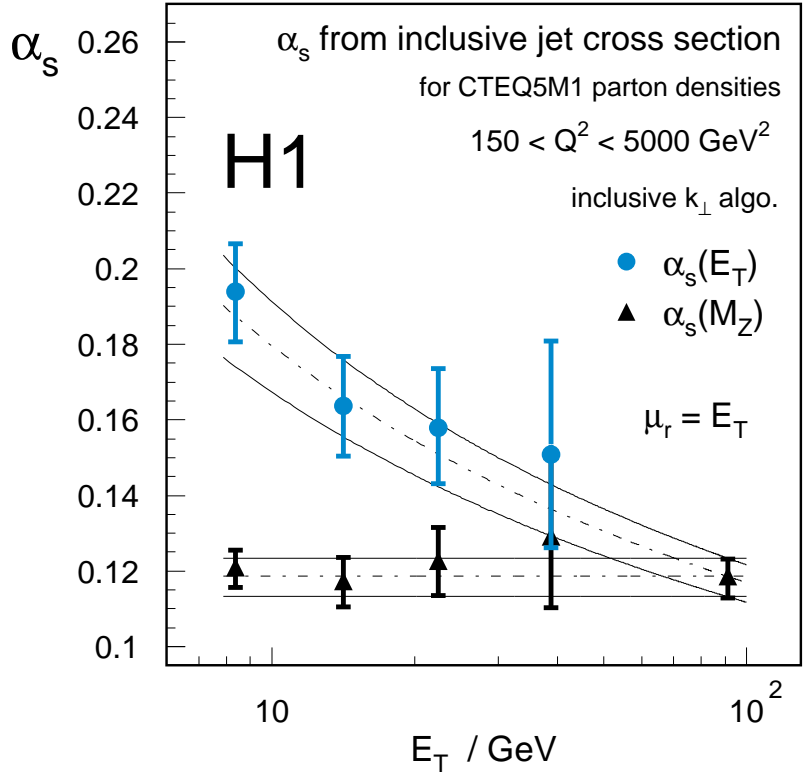,width=11cm}
\vskip-2mm
\caption{Determination of $\alps$ from the inclusive jet cross 
section using the inclusive $k_\perp$ algorithm
for the renormalization scale $\mu_r = E_T$.
The results are shown for each $E_T$ value (circles)
including experimental and theoretical uncertainties.
The single values are extrapolated to the $Z^0$-mass (triangles).
The final result for $\alpsmz$ (rightmost triangle) is obtained in a 
combined fit.
The lower curves represent the combined fit result and its
uncertainties and the upper curves indicate the prediction of the 
renormalization group equation for its energy evolution.}
\label{fig:as_result}
\end{figure}

Having checked that the data are consistent over the whole
range of $Q^2$ and $E_T$ combined fits are made to 
groups of data points.
To study the $E_T$ dependence of $\alpset$ the four data points 
of the same $E_T$ at different $Q^2$ are combined and four 
values of  $\alpset$ are extracted.
The results are shown in Fig.~\ref{fig:as_result}.
The four single values are evolved to $\alpha_s(M_Z)$.
A combined fit to all 16 data points gives $\cndof = 3.80 / 15$ 
which is rather small, possibly reflecting a conservative 
estimate of systematic uncertainties.
The central result is
\begin{equation}  
\alpha_s(M_Z) \, = \, 0.1186 \, 
 \pm 0.0030 \, ({\rm exp.}) 
\, ^{+0.0039}_{-0.0045} \, ({\rm th.})  \hskip20mm (\mu_r = E_T)  \, ,
\label{eq:as_et}
\end{equation}
in good agreement with the current world average of
$\alpsmz = 0.1184 \pm 0.0031$~\cite{bethke99}.
The statistical uncertainty of the result is very 
small ($\pm0.0007$).
The largest contribution to the experimental uncertainty 
comes from the hadronic energy scale of the LAr calorimeter.
The theoretical uncertainty  includes equal contributions from
the uncertainties of the hadronization corrections and 
the renormalization scale dependence.
The contribution due to the uncertainty of the parton distributions 
is  discussed below.

\subsubsection*{\boldmath Choice of $\sqrt{Q^2}$ as Renormalization 
Scale \label{sec:as_q}}

Another possible choice of the renormalization scale in the 
theoretical calculation is the four-momentum transfer $\sqrt{Q^2}$.
Analogous to the procedure applied before, an $\alpha_s$ 
determination is made for the renormalization scale $\mu_r = \sqrt{Q^2}$.
A combined fit to the 16 data points gives $\cndof = 3.87 / 15$
and a result
\begin{equation}
\alpha_s(M_Z) \, = \, 0.1227  ^{+0.0033}_{-0.0034} \, ({\rm exp.}) 
\, ^{+0.0055}_{-0.0060} \, ({\rm th.})  \hskip20mm (\mu_r = \sqrt{Q^2}) \, .
\label{eq:as_q}
\end{equation}
Comparing this result with the one obtained for $\mu_r = E_T$ 
in (\ref{eq:as_et}), the central value is seen too be shifted 
by $+0.0041$ and the theoretical uncertainty to have
increased substantially.
This is due to the stronger renormalization scale dependence
in the perturbative cross sections for $\mu_r = \sqrt{Q^2}$
compared to $\mu_r =E_T$.
Within the increased uncertainty contribution from the 
renormalization scale dependence for $\mu_r = \sqrt{Q^2}$
both results are consistent.

\subsubsection*{Using Different Parameterizations of Parton Distributions}

The central fit results are obtained for the parton distributions 
from the CTEQ5M1 parame\-terization~\cite{cteq5}.
The QCD fits are repeated using all parameterizations
from recent global fits which have been performed in 
next-to-leading logarithmic accuracy in the 
$\overline{\rm MS}$ scheme.
These include all sets from the fits
CTEQ5~\cite{cteq5}, Botje99~\cite{botje99}, MRST99~\cite{mrst99}, 
 CTEQ4~\cite{cteq4}, MRSR~\cite{mrsr}, 
\mbox{MRSAp~\cite{mrsap}} and the sets from the gluon 
uncertainty study~\cite{cteqgluon} by the 
CTEQ collaboration\footnote{The parameterization from 
GRV98~\cite{grv98} can not be used since the parameterizations of 
the charm and the bottom quark densities are not provided.}.
Many of these fits have provided sets of parton distributions for 
different assumptions for $\alpsmz$.
Using these sets of parton distributions, the dependence of our 
results on the initially assumed $\alpsmz$ is studied.

\begin{figure}
\centering
 \epsfig{file=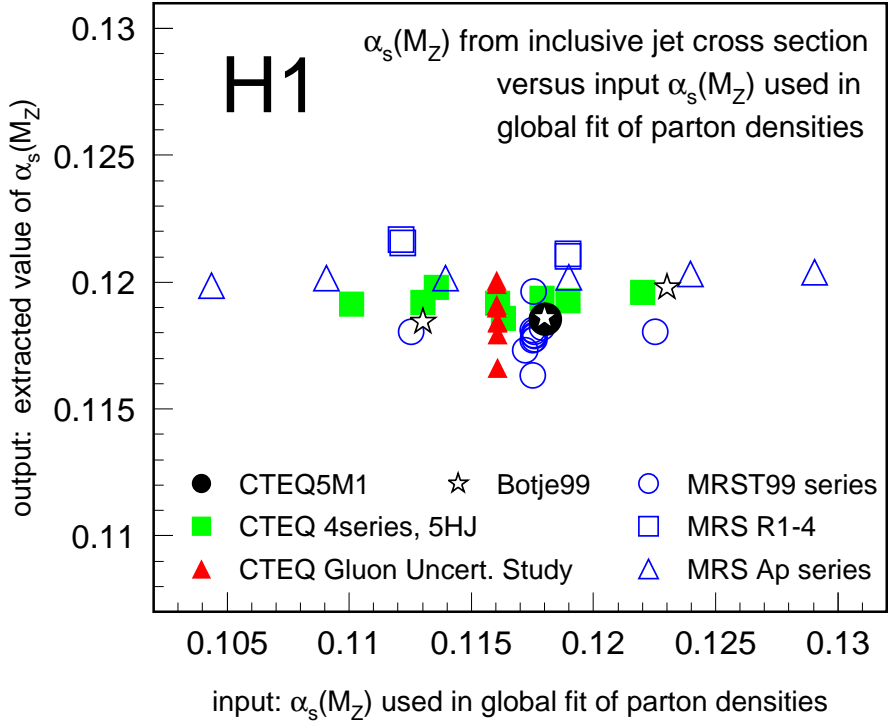,width=12.4cm}
\caption{Dependence of the $\alpha_s(M_Z)$ fit result 
(for $\mu_r = E_T$) on the
parton distributions used in the fit.
The results are displayed as a function of the
$\alpha_s(M_Z)$ value used in the corresponding global fit 
of the parton distributions.
The correlation is shown for a comprehensive collection of
different global fits.} 
\label{fig:pdf}
\end{figure}

The $\alpsmz$ results obtained for the different parton 
distributions are shown in Fig.~\ref{fig:pdf} as a function of 
the $\alpha_s(M_Z)$ value used in the corresponding global fit.
The range of the variations of the result is small and
no significant correlation to the initially assumed $\alpsmz$ 
is seen.
The largest deviations from the central result 
given in (\ref{eq:as_et}) 
are obtained with the MRSR3 parameterization ($+0.0031$) 
and for the set MRST99(g$\downarrow$) ($-0.0022$).
Using the central parameterizations from the most recent 
analyses, results of
$\alpsmz=0.1179$ for MRST99 and 
$\alpsmz=0.1186$ for Botje99 
are obtained which are very close to the result
obtained for CTEQ5M1.

\subsubsection*{Uncertainties in the Parton Distributions}

A determination of the uncertainties of parton density 
functions (pdfs) has only recently become available~\cite{botje99}.
This makes it possible to propagate these uncertainties into the 
predictions of physical quantities.
While earlier attempts were restricted to a limited number of 
variations of single parton flavors~\cite{mrst99,cteqgluon}
the fit performed by Botje~\cite{botje99} does not only
provide para\-meterizations of the central results, but also 
the covariance matrix $V_{ij}$ of the 28 fit parameters $p_i$
used, including
the  statistical and experimental systematic uncertainties.
In addition further systematic studies were performed
in~\cite{botje99}
by repeating the fit under different physical assumptions;
the corresponding deviations are, however, not included in the 
covariance matrix, but presented as single results.
The combined information is used to determine the uncertainty
of the $\alpsmz$ fit result by computing the contributions
from the covariance matrix and the single systematic studies
and add their contributions in quadrature.
The uncertainty from the parton density functions is then 
given by
\begin{eqnarray}  
\Delta^{\rm pdf} \alpsmz &  =  &
\sqrt{ \sum_{i,j} \frac{ \partial \alpsmz}{\partial p_i} \, V_{ij} \,
\frac{ \partial \alpsmz}{\partial p_j}}
  \; \oplus \;  
\sqrt{\sum_k \left( 
(\Delta\alpsmz)_k^{\rm syst.}
\right)^2 \; }         \nonumber \\ 
 &  =  & \hskip5.5mm
\pm0.0019 \: (\mbox{pdf: stat.\ \& exp.}) \; \oplus 
\hskip6mm 
^{+0.0027}_{-0.0013} \: (\mbox{pdf: fit syst.})   \nonumber \\  
& = & \hskip5.5mm  ^{+0.0033}_{-0.0023} \: (\mbox{\rm pdf})   \; .
\label{eq:as_pdf_unc}
\end{eqnarray}
The largest single contribution comes from the factorization
scale dependence which accounts for $^{+0.0020}_{-0.0003}$ 
in $\Delta \alpsmz$.
In fact, the uncertainty from the parton density functions, 
determined using this procedure is slightly larger than 
the spread observed in Fig.~\ref{fig:pdf}.
The value from (\ref{eq:as_pdf_unc}) is taken as the 
uncertainty of our $\alpsmz$ result due to the 
parton density functions.
The final result is then 
\begin{equation}  
\alpha_s(M_Z) \, = \, 0.1186 \, 
 \pm 0.0030 \, ({\rm exp.}) 
\, ^{+0.0039}_{-0.0045} \, ({\rm th.})
\, ^{+0.0033}_{-0.0023} \, ({\rm pdf})  
 \hskip10mm (\mu_r = E_T)  \, .
\label{eq:as_et_pdf}
\end{equation}

\subsubsection*{Fits to Other Observables}

\begin{figure}
\centering
 \epsfig{file=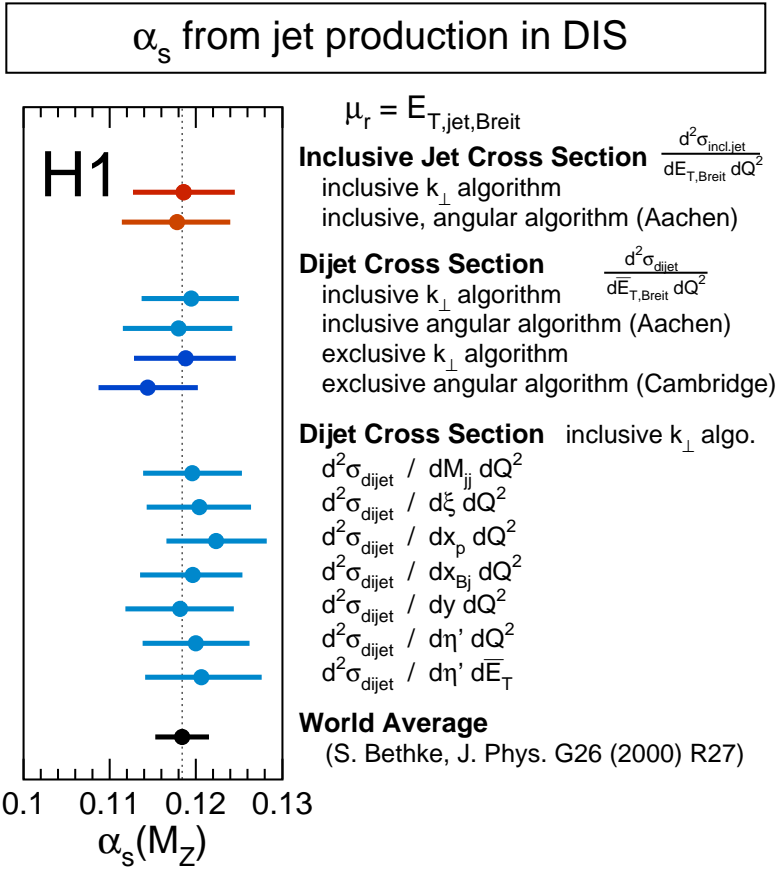,width=10.4cm}
\vskip-1mm
\caption{Comparison of $\alpsmz$ results from fits to different
double-differential jet distributions.}
\label{fig:as_allalgo}
\label{fig:as_dijvar}
\end{figure}

To test the stability of the central fit result the same QCD fits 
are made to some of the other jet distributions measured.
Included are fits to the 
differential inclusive jet cross section              \linebreak[4]
${\rm d}^2 \sigma_{\rm jet} /{\rm d}E_T {\rm d}Q^2$
using the Aachen algorithm and
the double-differential dijet cross section as
a function of various variables
for all four jet algorithms mentioned.
For the latter the renormalization scale is chosen
to be $\mu_r = \avet$.
The results of $\alpsmz$ from these fits 
including experimental, theoretical and the pdf uncertainties
are displayed in Fig.~\ref{fig:as_allalgo}.
All $\alpsmz$ values are in good agreement with each other,
with the central fit result given in (\ref{eq:as_et_pdf}) 
and with the current world average value.
The results for the exclusive jet algorithms have larger 
theoretical uncertainties due to the larger hadronization 
corrections.

\subsection{Determination of the Gluon and the Quark Densities 
in the Proton\label{sec:quarkgluon}}

The measurement of $\alpha_s$ described in section~\ref{sec:fitalps} 
depends on external knowledge of the parton content of the proton, 
and in particular on the uncertainty in the pdfs of the proton. 
The determined value of $\alpha_s$ is found to be consistent 
with measurements in which no initial state hadrons are involved,
for example in $e^+e^-$ annihilation to hadrons~\cite{jadeopal}. 
The validity of pQCD at NLO in jet production in DIS is thereby 
demonstrated unequivocally to within the accuracy with which the 
strong coupling constant $\alpha_s$ is known.

It is therefore appropriate to pursue a determination of the 
parton density functions of the proton in NLO pQCD using 
measurements of jet production cross sections in DIS, assuming 
the value of the strong coupling constant $\alpha_s$ from 
external measurements. 
Such a determination is important for two reasons.
First the measurement is in principle sensitive directly to both quark
and gluon content in the proton, in contrast with studies of the
evolution in $x_{\rm Bj}$ and $Q^2$ of the proton structure function $F_2$
where there is only direct sensitivity to the quark content. 
Second the range in the fractional momentum variable $\xi$ 
covered by a measurement using jets is different from that 
attained with $F_2$ measurements.

In the second step of the QCD analysis the sensitivity
of the jet cross sections to the gluon density in the proton
is exploited.
The dijet cross section as a function of $\xi$ is directly
sensitive to the gluon density at $x=\xi$.
The inclusion of the inclusive jet cross section as a function
of $E_T$ maximizes the accessible range in $x$.
Data from a recent measurement of the inclusive DIS cross 
section~\cite{h1highq2} give strong, direct constraints
on the quark density $x\Delta(x)$.
Furthermore the strong coupling constant is set to the world 
average value $\alpsmz = 0.1184 \pm 0.0031$~\cite{bethke99}.

In the central fit 
the dijet cross section 
${\rm d}\sigma^2_{\rm dijet} / {\rm d}\xi {\rm d}Q^2$
at $150 < Q^2 < 5\,000\GeV^2$,
the inclusive jet cross section
${\rm d}\sigma^2_{\rm jet} / {\rm d}E_T {\rm d}Q^2$
at $150 < Q^2 < 5\,000\GeV^2$
and the reduced inclusive DIS cross section 
$\tilde{\sigma}(\xbj,Q^2)$ from~\cite{h1highq2}
in the range $150 \le Q^2 \le 1\,000\GeV^2$ 
($0.032 < \xbj < 0.65$)
are used.
 The gluon density and the $x\Delta(x)$ quark density are 
 parameterized  by
\be
xP(x)  = A \; x^b \; (1-x)^c \; (1+dx)      \label{eq:pdfpara}
\end{equation}
where $xP(x)$ stands for $xG(x)$ or $x\Delta(x)$.

\begin{figure}
\centering
 \epsfig{file=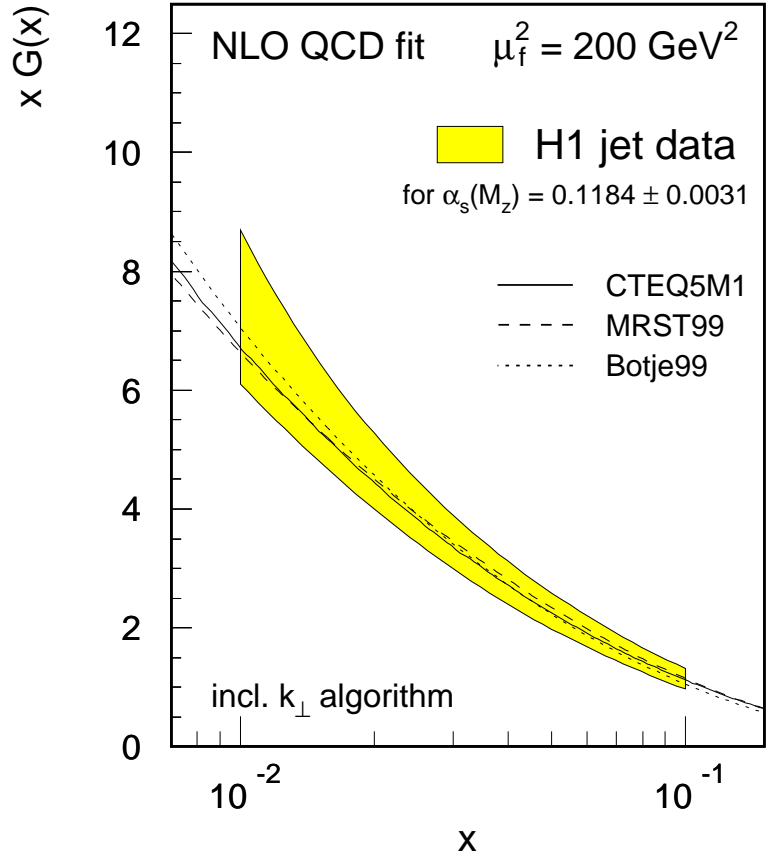,width=10.7cm}
\caption{The gluon density $xG(x)$  
in the proton, determined in a combined QCD fit
to the inclusive DIS cross section, the inclusive 
jet cross section and the dijet cross section. 
The jet cross sections are measured using the inclusive 
$\kperp$ jet algorithm.
The error band includes the experimental and the theoretical 
uncertainties as well as the uncertainty of $\alpsmz$.}
\label{fig:gluon}
\end{figure}

The gluon density is determined in the range $0.01 < x < 0.1$
at the factorization scale $\mu_f = \sqrt{200}\GeV$
with $\cndof = 61.16/105$.
The result is shown in Fig.~\ref{fig:gluon}.
Displayed is the error band, including all experimental and
theoretical uncertainties and the uncertainty from the
value of the world average value of $\alpsmz$.
The result is  seen to be in good agreement with
results from recent global data analyses.
The integral of the gluon density over the range $0.01 < x < 0.1$ 
has been determined to be
\bea
\int_{0.01}^{0.1} {\rm d}x \; x\,G(x,\mu_f^2=200\GeV^2) & = &
0.229 \;^{+0.031}_{-0.030}\mbox{(tot.)} \, , \\
 & = & 0.229 
\;^{+0.016}_{-0.015} \mbox{(exp.)} 
\;^{+0.019}_{-0.021} \mbox{(th.)}  
\;^{+0.018}_{-0.015} (\Delta \alps)  \, . \nonumber
\eea
This means that at the scale $\mu_f = \sqrt{200}\GeV$
gluons with a momentum fraction in the range $0.01 < x < 0.1$
carry 23\% of the total proton momentum.
This result is in good agreement with the results from global 
fits for which the integral has the values 
\begin{equation}
\mbox{CTEQ5M1:}\;\;  0.227 \; , \hskip4mm
\mbox{MRST99:}\;\;  0.232 \; , \hskip4mm
\mbox{GRV98HO:}\;\; 0.235 \; , \hskip4mm
\mbox{Botje99:}\;\; 0.227 \; .
\end{equation}

The quark density $x\Delta(x,\mu^2_f=200\GeV^2)$ 
determined in this fit is also close to results 
from global fits.
To test the stability of the results various cross checks
have been performed~\cite{wobisch2000a}:

\begin{itemize}

\item[(a)] Different parameterizations of the parton densities:
\begin{eqnarray}
\mbox{3 parameters\  \hskip7mm}&xP(x) 
       \; = & A \; x^b \; (1-x)^c    \nonumber \\
\mbox{5 parameters\  (I) \hskip7mm} 
      &xP(x) \; = & A \; x^b \; (1-x)^c \; (1+dx^e)
  \nonumber \\
\mbox{5 parameters\ (II) \hskip7mm} &xP(x) \; = & A \; x^b \; 
(1-x)^c \; (1+d\sqrt{x} + ex) 
   \nonumber 
\end{eqnarray}

While the central result has been obtained using the
4-parameter ansatz in (\ref{eq:pdfpara}),
the gluon density is unchanged when using other
parameterizations and the quark density is stable 
if at least four parameters are used.

\item[(b)]  Fits to subsets of the data:
The fit has been applied to  two subsamples of the data
with $Q^2<300\GeV^2$ and $Q^2>300\GeV^2$ and both, the gluon and
the quark results are unchanged.

\item[(c)]  Fits to other jet distributions:
The fits have been repeated using other jet distributions measured with
the inclusive $\kperp$ algorithm and also to jet distributions 
measured with other jet algorithms.
In all cases the results are consistent with each other.

\end{itemize}

\subsection{\boldmath Simultaneous Determination of $\alps$ and
the Proton pdfs}

\begin{figure}
\centering
 \epsfig{file=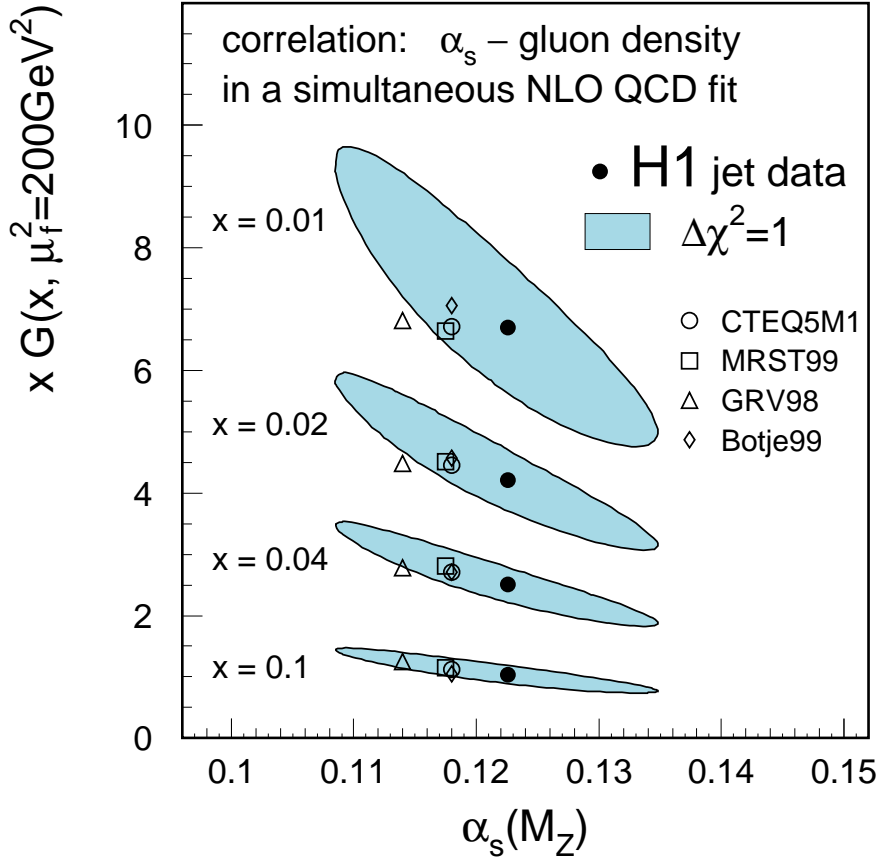,width=11.9cm}
\caption{The correlation of the fit results for $\alpsmz$ 
and the gluon density at four different values of $x$,
determined in a simultaneous QCD fit to the inclusive
DIS cross section, the inclusive jet cross section and 
the dijet cross section. 
The jet cross sections are measured using the inclusive 
$\kperp$ jet algorithm.
The central fit result is indicated by the full marker.
The error ellipses include the experimental and the 
theoretical uncertainties.}
\label{fig:asg_cor}
\end{figure}

In the above, $\alps$ or the gluon density are extracted using
external knowledge for the other.  
A more independent test of pQCD can
be made in a simultaneous determination of both quantities.
Such a determination has been performed by fitting 
the parton densities and $\alpsmz$
using the same data sets as in the previous section,
the inclusive DIS cross section, the inclusive jet cross section
${\rm d}^2 \sigma_{\rm jet} / {\rm d}E_T {\rm d}Q^2$ and the dijet cross 
section ${\rm d}^2 \sigma_{\rm dijet} / {\rm d}\xi {\rm d}Q^2$
(again measured with the inclusive $\kperp$ algorithm).
The gluon and the quark distributions are parameterized according to
the 4-parameter formula in (\ref{eq:pdfpara}).
The simultaneous fit yields $\cndof = 61.19/104$ and a result
for the quark distributions identical to that which is obtained in 
the fit with a constrained $\alpsmz$.
The results of this simultaneous fit are displayed in 
Fig.~\ref{fig:asg_cor} as a
correlation plot between $\alpsmz$ and the gluon density 
evaluated at four different values of $x=0.01,\,0.02,\,0.04,\,0.1$
which lie in the range where the jet cross sections are sensitive.
The central fit result is indicated by the full marker
and the error ellipse is the contour along which
the $\chi^2$ of the fit is by one larger than the minimum 
(including experimental and theoretical uncertainties).
The ellipticity of the contours 
indicate that the data included in this analysis
are very sensitive to the product
$\alps \cdot xg(x)$ but do not allow a determination
of both parameters simultaneously with useful precision.

Also included in Fig.~\ref{fig:asg_cor} are the results 
from global fits.
All of these results are within the error ellipses
except for GRV98~\cite{grv98} (at $x<0.04$) which uses
a relatively small value of $\alpsmz = 0.114$.

The stability of the results in  Fig.~\ref{fig:asg_cor} 
has been tested in a similar way as already described
in section~\ref{sec:quarkgluon}.
Fits have been performed excluding either the low ($ < 200\GeV^2$)
or the high ($>600\GeV^2$) $Q^2$ data.
Although the fits give consistent central results,
the high $Q^2$ data are needed to achieve a stable 
determination of the contour of the error
ellipsoid.

\section{Summary}

Jet production has been studied in the Breit frame in
deep-inelastic positron-proton collisions
at a center-of-mass energy
of $\sqrt{s}=300\GeV$.
In the range of four-momentum transfers 
$5 < Q^2 < 15\,000\GeV^2$
and transverse jet energies $7 < E_T < 60\GeV$ 
dijet and inclusive jet cross sections have been measured 
as a function of various variables
using $\kperp$ and angular ordered jet clustering algorithms.
Perturbative QCD in next-to-leading order in $\alps$
gives a good description of all observables
for which next-to-leading order corrections are not too large
and for which hadronization corrections are small.
For those observables with moderately large
hadronization corrections the deviations between data
and the perturbative calculations are always consistent
with the size of the hadronization corrections
as predicted by phenomenological models.
Only at $Q^2 < 10\GeV^2$ do
the theoretical predictions fail to describe the size
of the measured jet cross sections.
In this region, however, the NLO corrections are
large, with $k$-factors above two indicating that
NLO calculations are not reliable
and that it is likely that the perturbative predictions
receive large contributions from higher orders
in $\alps$ which can account for the observed 
difference.

QCD analyses of the data have been performed in
the region of $Q^2 > 150\GeV^2$,
where NLO calculations are reliable,
using the inclusive $\kperp$ jet algorithm for which 
hadronization corrections are smallest.
In a first step $\alps$ has been determined
in a fit to the inclusive jet cross section
as a function of the transverse jet energy.
Here the knowledge of the parton density functions 
of the proton is taken from 
the results of global fits.
The observed $E_T$ dependence of $\alps$ is consistent with 
the prediction of the renormalization group equation
and a combined fit to the data yields
\bea
\alpha_s(M_Z) & = & 0.1186 \, \pm \, 0.0059\, ({\rm tot.}) 
 \nonumber \\
    & = & 0.1186 \, 
 \pm 0.0030 \, ({\rm exp.}) 
\, ^{+0.0039}_{-0.0045} \, ({\rm th.})
\, ^{+0.0033}_{-0.0023} \, ({\rm pdf})   \, .  \nonumber 
\end{eqnarray}
This result is seen to be stable when the fit 
is performed to a variety of jet distributions
measured with different jet algorithms.

Including H1 data on the inclusive neutral current 
DIS cross section, the jet data have been used for
a consistent determination of the gluon density in the 
proton together with the quark densities.
Setting $\alps$ to the world average value~\cite{bethke99}
within its uncertainty of 
$\alpsmz = 0.1184 \pm 0.0031$
the gluon density is determined in the range
of momentum fractions $0.01 < x < 0.1$ at a 
factorization scale of the order of the transverse
jet energies $\mu_f = \sqrt{200}\GeV$ in the 
$\overline{\rm MS}$ scheme.
The integral over the range $0.01 < x < 0.1$ is 
determined to be
\bea
\int_{0.01}^{0.1} {\rm d}x \; x\,G(x,\mu_f^2=200\GeV^2) & = &
0.229 \;^{+0.031}_{-0.030}\mbox{(tot.)} \, , \nonumber \\
 & = & 0.229 
\;^{+0.016}_{-0.015} \mbox{(exp.)} 
\;^{+0.019}_{-0.021} \mbox{(th.)}  
\;^{+0.018}_{-0.015} (\Delta \alps)  \, . \nonumber
\eea
This result, as well as the differential distribution
in $x$, are in good agreement with the results obtained
in global fits.

Finally
$\alps$ and the gluon density in the proton 
have been determined simultaneously
using data with direct sensitivity to both.
The results and their uncertainties show a large 
anticorrelation.
Here the single results of $\alps$ and the gluon density
have relatively large uncertainties,
but the strong anticorrelation
of the combined result clearly demonstrates the high sensitivity
of the jet data to both.

\section*{Acknowledgments}
We wish to thank Erwin Mirkes, Dieter Zeppenfeld, Mike H. Seymour
and Bj\"orn P\"otter for many helpful discussions.
We are grateful to the HERA machine group whose outstanding efforts
have made and continue to make this experiment possible.
We thank the engineers and technicians for their work in constructing and now 
maintaining the H1 detector, our funding agencies for financial support,
the DESY technical staff for continual assistance, and the DESY directorate
for the hospitality which they extend to the non-DESY members of the
collaboration.


\newpage
\begin{appendix}

\section*{Tables of Experimental Results}


In the following those jet cross sections are listed
which have been used in the QCD analyses to obtain 
the central results.
The numbers of
further distributions can be found in~\cite{wobisch2000a}
or are available on request from the H1 collaboration.

\subsubsection*{\boldmath The Inclusive Jet Cross Section 
${\rm d}^2\sigma_{\rm jet} / ({\rm d}E_T \, {\rm d}Q^2)$}

\begin{center}
\scriptsize \sf
\begin{tabular}{|c||l|}
\hline
bin number &  corresponding $Q^2$ range \\
\hline
\,\, 1 \,\, & $150<Q^2<200\GeV^2$ \\
\,\, 2 \,\, & $200<Q^2<300\GeV^2$ \\
\,\, 3 \,\, & $300<Q^2<600\GeV^2$ \\
\,\, 4 \,\, & $600<Q^2<5000\GeV^2$ \\
\hline
\end{tabular}
\hskip20mm
\begin{tabular}{|c||l|}
\hline
bin letter &  corresponding $E_T$ range \\
\hline
\,\, a \,\, &  $ \;\,7 < E_T < 11\GeV$ \\ 
\,\, b \,\, &  $ 11 < E_T < 18\GeV$ \\ 
\,\, c \,\, &  $ 18 < E_T < 30\GeV$ \\ 
\,\, d \,\, &  $ 30 < E_T < 50\GeV$ \\ 
\hline
\end{tabular}
\end{center}

\begin{table}[h!]
\tiny \sf
\centering
\begin{tabular}{|r||r|r ||r|r ||r|r| r|r ||r |r |r|r |r||r|} 
\hline
\multicolumn{15}{|c|}{ } \\
\multicolumn{15}{|c|}{\normalsize the inclusive jet cross section 
\, ${\rm d}^2\sigma_{\rm jet} / ({\rm d}E_T \, {\rm d}Q^2)$ \, --- \,
inclusive $k_\perp$ jet algorithm } \\
\multicolumn{15}{|c|}{ } \\
\hline
 & & &\multicolumn{2}{c||}{}  & \multicolumn{2}{c|}{} &\multicolumn{2}{c||}{} 
  & 
\multicolumn{5}{c||}{\underline{single contributions to correlated uncertainty}}
  &\\
bin & cross & statistical & \multicolumn{2}{c||}{total} & 
\multicolumn{2}{c|}{uncorrelated} &\multicolumn{2}{c||}{correlated}  & 
model dep. & positron  & positron & \multicolumn{2}{c||}{LAr hadr.} 
& hadroniz. \\
No. & section & uncert. &\multicolumn{2}{c||}{uncertainty}  & 
\multicolumn{2}{c|}{uncertainty} & \multicolumn{2}{c||}{uncertainty} 
& detector corr. & energy scale & polar angle & 
\multicolumn{2}{c||}{energy scale} &  correct. \\
  & (in pb) & (in percent) &\multicolumn{2}{c||}{(in percent) }  & 
\multicolumn{2}{c|}{(in percent) } & \multicolumn{2}{c||}{(in percent) } 
& (in percent) & (in percent) & (in percent) & 
\multicolumn{2}{c||}{(in percent)} &  (percent) \\
\hline
\hline
   1 a&   62.220&$\pm$  2.9&  9.6& -9.5&  7.6&  -7.4&  5.9&  -5.9&$\pm$  4.7&  0.8&  1.7&  2.6& -2.6&  -7.8\\
   1 b&   26.084&$\pm$  4.4& 16.3&-13.9& 13.6& -11.3&  9.0&  -8.0&$\pm$  6.3&  0.8&  1.6&  6.0& -4.4&  -5.0\\
   1 c&    5.819&$\pm$  9.2& 13.8&-15.6& 12.3& -13.8&  6.3&  -7.2&$\pm$  4.3&  1.0&  1.7&  3.9& -5.3&  -4.5\\
   1 d&    0.719&$\pm$ 27.8& 34.0&-35.4& 32.4& -33.4& 10.4& -11.6&$\pm$  3.4&  1.9&  2.7&  9.2&-10.5&  -4.7\\
\hline
   2 a&   62.256&$\pm$  2.6&  7.6& -7.2&  6.4&  -6.1&  4.1&  -3.9&$\pm$  2.3&  0.2&  1.0&  2.8& -2.5&  -8.1\\
   2 b&   29.802&$\pm$  3.7& 12.2&-12.7& 10.3& -10.8&  6.5&  -6.7&$\pm$  4.0&  0.7&  0.2&  4.9& -5.2&  -4.5\\
   2 c&    6.989&$\pm$  7.6& 16.8&-14.8& 14.3& -12.4&  8.9&  -8.1&$\pm$  6.7&  0.3&  1.9&  5.3& -3.8&  -4.6\\
   2 d&    0.994&$\pm$ 20.0& 28.7&-33.5& 26.0& -29.9& 12.1& -15.0&$\pm$  8.6&  2.7&  1.9&  7.7&-11.8&  -4.9\\
\hline
   3 a&   61.577&$\pm$  2.7&  5.8& -6.1&  4.9&  -5.3&  3.0&  -3.1&$\pm$  1.2&  0.7&  0.8&  2.0& -2.2&  -8.1\\
   3 b&   35.010&$\pm$  3.5& 11.9& -9.9& 10.1&  -8.2&  6.3&  -5.5&$\pm$  3.5&  0.5&  1.4&  4.8& -3.6&  -4.0\\
   3 c&    9.644&$\pm$  6.6& 15.7&-17.9& 13.3& -15.3&  8.4&  -9.4&$\pm$  4.2&  1.6&  3.7&  5.9& -7.2&  -4.5\\
   3 d&    1.362&$\pm$ 20.0& 36.3&-29.7& 32.1& -26.4& 17.1& -13.7&$\pm$ 11.4&  1.1&  1.6& 12.5& -7.1&  -4.6\\
\hline
   4 a&   46.515&$\pm$  3.1&  7.4& -6.8&  6.3&  -5.8&  3.8&  -3.5&$\pm$  0.7&  0.0&  1.4&  3.1& -2.7&  -9.0\\
   4 b&   26.409&$\pm$  4.1&  8.8& -8.8&  7.6&  -7.6&  4.4&  -4.4&$\pm$  1.6&  0.6&  1.5&  3.5& -3.5&  -4.0\\
   4 c&   11.288&$\pm$  6.0& 11.5&-11.6& 10.3& -10.3&  5.3&  -5.4&$\pm$  2.4&  0.3&  0.5&  4.4& -4.6&  -3.2\\
   4 d&    1.993&$\pm$ 15.1& 27.2&-23.0& 24.6& -21.2& 11.5&  -8.9&$\pm$  1.2&  0.9&  1.9& 11.1& -8.4&  -3.2\\
\hline
\end{tabular}
\rm \normalsize
\caption{Results of the inclusive jet cross section measurement
 using the inclusive $k_\perp$ jet algorithm.
The listing includes all experimental uncertainties 
(as described in section~\ref{sec:expunc})
which are here separated into the correlated and the 
uncorrelated part.
Since the interpretation of the results 
(as e.g.\  in a QCD analysis) does not require
the knowledge of the separate contributions to the 
uncorrelated part of the uncertainties,
only the total uncorrelated uncertainty is presented
while the single contributions to the correlated
uncertainty are listed in extra columns for all sources.
The uncertainty from the hadronic energy scale
of the Liquid Argon calorimeter is quoted asymmetrically.
The left (right) value corresponds to an increase
(decrease) of the calibration constants.
The uncertainties of the positron energy and the positron 
polar angle are defined to be symmetric
by taking the maximum of the upwards and downwards deviations.
The signs are quoted for a positive variation
of the corresponding source.
Note that only the {\em correlated} contribution
from these sources is listed.
As described in section~\ref{sec:expunc}
some of these sources contribute also to the 
uncorrelated uncertainty.
The latter contribution is already contained in the 
(quadratic) sum of all uncorrelated uncertainties.
The total correlated uncertainty includes also 
the contribution of $\pm 1.5\%$ from the uncertainty in
the determination of the luminosity.
In the right column we have also included the 
size of the hadronization corrections as determined 
by the procedure described in section~\ref{sec:fittec}.
\label{tab:res_incl_ik}}
\end{table}

\clearpage  
\subsubsection*{\boldmath The Dijet Cross Section 
$d^2\sigma_{\rm dijet} / (d\xi \, dQ^2)$}

\begin{table}[h!]
\scriptsize \sf
\centering

 \begin{tabular}{cc}

 \begin{tabular}[t]{|c||c||c|}
\hline
bin No. &  corresponding $Q^2$ range & $\xi $ range \\
\hline
 1 a & $5 < Q^2 < 10\GeV^2$ &
 0.004 $< \xi <$  0.01   \\
 1 b & &
 0.01 $< \xi <$  0.025   \\
 1 c & &
 0.025 $< \xi <$  0.05   \\
 1 d & &
 0.05 $< \xi <$  0.1   \\
 \hline
 2 a & $10 < Q^2 < 20\GeV^2$ &
 0.004 $< \xi <$  0.01   \\
 2 b & &
 0.01 $< \xi <$  0.025   \\
 2 c & &
 0.025 $< \xi <$  0.05   \\
 2 d & &
 0.05 $< \xi <$  0.1   \\
 \hline
 3 a & $ 20 < Q^2 < 35\GeV^2$ &
 0.004 $< \xi <$  0.01   \\
 3 b & &
 0.01 $< \xi <$  0.025   \\
 3 c & &
 0.025 $< \xi <$  0.05   \\
 3 d & &
 0.05 $< \xi <$  0.1   \\
 \hline
 4 a & $ 35 < Q^2 < 70\GeV^2$ &
 $0.004 < \xi < 0.01$   \\
 4 b & &
 $0.01 < \xi <  0.025 $  \\
 4 c & &
 $0.025 < \xi < 0.05$   \\
 4 d & &
 $0.05 <\xi < 0.1$   \\
 \hline
 \end{tabular}
&
\begin{tabular}[t]{|c||c||c|}
\hline
bin No. &  corresponding $Q^2$ range & $\xi $ range \\
\hline
 5 a  & $150<Q^2<200\GeV^2$ & $0.009< \xi < 0.017$  \\
 5 b  & $ $ & $0.017 < \xi < 0.025$  \\
 5 c  & $ $ & $0.025< \xi <0.035 $  \\
 5 d  & $ $ & $0.035< \xi <0.05 $  \\
 5 e  & $ $ & $0.05< \xi <0.12 $  \\
\hline
 6 a  & $200<Q^2<300\GeV^2$ & $0.01< \xi <0.02 $  \\
 6 b  & $ $ & $0.02< \xi <0.03 $  \\
 6 c  & $ $ & $0.03< \xi <0.04 $  \\
 6 d  & $ $ & $0.04< \xi <0.06 $  \\
 6 e  & $ $ & $0.06< \xi <0.15 $  \\
\hline
 7 a  & $300<Q^2<600\GeV^2$ & $0.015 < \xi < 0.025$  \\
 7 b  & $ $ & $0.025< \xi < 0.035$  \\
 7 c  & $ $ & $0.035< \xi <0.045 $  \\
 7 d  & $ $ & $0.045< \xi <0.065 $  \\
 7 e  & $ $ & $0.065< \xi <0.18 $  \\
\hline
 8 a  & $600<Q^2<5000\GeV^2$ & $0.025 < \xi <0.045 $  \\
 8 b  & $ $ & $0.045< \xi <0.065 $  \\
 8 c  & $ $ & $0.065< \xi < 0.1 $ \\
 8 d  & $ $ & $0.1< \xi < 0.3 $  \\
\hline
\end{tabular}

\end{tabular}
\end{table}

\begin{table}[h!]
\tiny \sf
\centering
\begin{tabular}{|r||r|r ||r|r ||r|r| r|r ||r |r |r|r |r||r|} 
\hline
\multicolumn{15}{|c|}{ } \\
\multicolumn{15}{|c|}{\normalsize the dijet cross section 
\, $d^2\sigma_{\rm dijet} / (d\xi \, dQ^2)$ \, --- \,
inclusive $k_\perp$ jet algorithm} \\
\multicolumn{15}{|c|}{ } \\
\hline
 & & &\multicolumn{2}{c||}{}  & \multicolumn{2}{c|}{} &\multicolumn{2}{c||}{} 
  & 
\multicolumn{5}{c||}{\underline{single contributions to correlated uncertainty}}
  &\\
bin & cross & statistical & \multicolumn{2}{c||}{total} & 
\multicolumn{2}{c|}{uncorrelated} &\multicolumn{2}{c||}{correlated}  & 
model dep. & positron  & positron & \multicolumn{2}{c||}{LAr hadr.} 
& hadroniz. \\
No. & section & uncert. &\multicolumn{2}{c||}{uncertainty}  & 
\multicolumn{2}{c|}{uncertainty} & \multicolumn{2}{c||}{uncertainty} 
& detector corr. & energy scale & polar angle & 
\multicolumn{2}{c||}{energy scale} &  correct. \\
  & (in pb) & (in percent) &\multicolumn{2}{c||}{(in percent) }  & 
\multicolumn{2}{c|}{(in percent) } & \multicolumn{2}{c||}{(in percent) } 
& (in percent) & (in percent) & (in percent) & 
\multicolumn{2}{c||}{(in percent)} &  (percent) \\
\hline
\hline
 1 a &     19.34 &  $\pm$      6.8 &   10.6 &  -10.0 &    8.5 &   -8.5 &    6.3 
&   -5.3 &  $\pm$  2.4 &   -0.2 &   -3.6 &    4.9 &   -3.0 &  -6.1  \\
 1 b &     83.84 &  $\pm$      3.8 &   10.8 &   -9.2 &    8.4 &   -6.3 &    6.8 
&   -6.7 &  $\pm$  1.1 &    2.7 &    4.8 &    3.7 &   -3.6 &  -9.3  \\
 1 c &     47.96 &  $\pm$      5.2 &    6.3 &  -10.1 &    4.8 &   -8.2 &    4.2 
&   -5.9 &  $\pm$  3.8 &    0.6 &    2.7 &    2.7 &   -3.5 &  -5.9  \\
 1 d &     18.72 &  $\pm$      7.4 &   16.8 &   -8.6 &   13.6 &   -6.9 &    9.9 
&   -5.2 &  $\pm$   6.2 &    2.6 &    2.1 &    6.8 &   -4.0 &  -9.3  \\
\hline
 2 a &     14.67 &  $\pm$      7.1 &    8.6 &   -9.2 &    6.9 &   -7.6 &    5.1 
&   -5.2 &  $\pm$  2.2 &    1.6 &   -2.3 &    4.0 &   -3.8 &  -13.2  \\
 2 b &     66.71 &  $\pm$      3.6 &    8.0 &   -9.9 &    6.6 &   -8.7 &    4.5 
&   -4.7 &  $\pm$  2.0 &    1.0 &   -1.6 &    3.8 &   -3.8 &  -6.4  \\
 2 c &     39.42 &  $\pm$      5.0 &   12.2 &   -7.8 &   10.2 &   -5.6 &    6.6 
&   -5.4 &  $\pm$  3.2 &    1.1 &    3.2 &    5.4 &   -2.7 &  -5.6  \\
 2 d &     14.58 &  $\pm$      7.8 &   11.2 &  -13.3 &    8.8 &  -10.8 &    6.9 
&   -7.7 &  $\pm$  3.7 &   -1.7 &    4.0 &    5.1 &   -5.2 &  -11.4  \\
\hline
 3 a &      9.45 &  $\pm$      8.8 &   15.7 &  -11.9 &   13.4 &  -10.6 &    8.2 
&   -5.6 &  $\pm$  1.7 &    2.4 &    1.1 &    7.6 &   -4.6 &  -12.7  \\
 3 b &     49.42 &  $\pm$      4.0 &    7.9 &   -9.2 &    6.5 &   -8.1 &    4.5 
&   -4.4 &  $\pm$  1.1 &    0.8 &    1.7 &    3.8 &   -3.8 &  -7.0  \\
 3 c &     28.83 &  $\pm$      5.3 &    7.1 &  -11.4 &    5.4 &   -9.9 &    4.6 
&   -5.7 &  $\pm$  0.3 &   -1.8 &   -2.4 &    3.1 &   -4.8 &  -6.6  \\
 3 d &     10.90 &  $\pm$      9.6 &   12.4 &  -12.4 &   10.2 &   -9.8 &    7.1 
&   -7.6 &  $\pm$  4.7 &    3.1 &    1.8 &    5.9 &   -4.8 &  4.2  \\
\hline
 4 a &      6.46 &  $\pm$     10.6 &   16.7 &   -9.3 &   15.0 &   -7.6 &    7.4 
&   -5.3 &  $\pm$  1.5 &    1.9 &    1.9 &    6.7 &   -4.3 &  -3.0  \\
 4 b &     47.92 &  $\pm$      3.8 &    9.5 &   -7.0 &    8.2 &   -5.7 &    4.9 
&   -4.1 &  $\pm$   1.0 &   -1.3 &    2.1 &    3.8 &   -3.3 &  -4.7  \\
 4 c &     31.09 &  $\pm$      4.9 &    6.7 &   -9.8 &    5.4 &   -8.3 &    3.8 
&   -5.3 &  $\pm$  2.4 &   -1.1 &    1.1 &    3.1 &   -4.4 &  -5.0  \\
 4 d &     12.06 &  $\pm$      8.0 &   13.8 &   -9.0 &   11.9 &   -7.5 &    7.1 
&   -5.0 &  $\pm$  0.4 &    0.9 &    2.4 &    6.5 &   -4.3 &  -11.5  \\
\hline
   5 a&    4.147&$\pm$ 11.0& 24.7&-27.7& 19.9&-22.9& 14.7&-15.6&$\pm$ 13.7&  1.9&  1.8&  4.3& -6.9&  -6.0\\
   5 b&    6.272&$\pm$  9.1& 14.9&-13.8& 13.2&-12.2&  6.9& -6.4&$\pm$  3.4&  2.1&  2.8&  4.7& -3.8&  -5.3\\
   5 c&    6.544&$\pm$  9.8& 12.3&-13.2& 11.2&-12.0&  5.1& -5.6&$\pm$  3.1&  3.0&  1.2&  1.8& -3.0&  -5.7\\
   5 d&    5.059&$\pm$ 10.9& 14.3&-15.3& 13.2&-14.1&  5.3& -5.9&$\pm$  1.9&  2.3&  2.2&  3.5& -4.4&  -5.4\\
   5 e&    4.800&$\pm$ 11.3& 18.4&-17.8& 15.1&-14.5& 10.6&-10.3&$\pm$  7.9&  3.3&  5.3&  2.9& -1.7&  -6.2\\
\hline
   6 a&    6.324&$\pm$  8.8& 13.7&-18.1& 12.0&-16.0&  6.5& -8.6&$\pm$  3.8&  3.6&  1.2&  3.4& -6.5&  -6.1\\
   6 b&    7.309&$\pm$  7.8& 10.9&-12.5& 10.0&-11.3&  4.4& -5.3&$\pm$  2.6&  0.3&  1.0&  3.1& -4.3&  -4.8\\
   6 c&    6.023&$\pm$  8.6& 11.7&-11.8& 10.5&-10.5&  5.2& -5.5&$\pm$  1.1&  1.5&  3.7&  2.8& -3.2&  -5.0\\
   6 d&    5.512&$\pm$  8.9& 14.9&-11.9& 13.4&-11.0&  6.6& -4.6&$\pm$  2.6&  1.6&  1.9&  5.3& -2.4&  -4.8\\
   6 e&    4.186&$\pm$ 10.3& 17.8&-19.3& 14.7&-16.1& 10.1&-10.8&$\pm$  8.3&  3.1&  3.6&  3.0& -4.8&  -7.5\\
\hline
   7 a&    5.997&$\pm$  9.3& 12.7&-15.4& 11.9&-14.4&  4.5& -5.6&$\pm$  1.5&  0.9&  1.0&  3.8& -5.1&  -5.4\\
   7 b&    7.006&$\pm$  8.3& 11.3&-12.4& 10.1&-11.1&  5.0& -5.6&$\pm$  3.5&  0.4&  2.1&  2.4& -3.4&  -4.8\\
   7 c&    6.104&$\pm$  9.0& 16.0&-11.9& 14.5&-11.1&  6.9& -4.5&$\pm$  1.2&  2.2&  1.0&  6.1& -3.3&  -4.5\\
   7 d&    7.249&$\pm$  8.1& 10.1&-10.4&  9.0& -9.2&  4.7& -4.9&$\pm$  1.9&  0.6&  3.8&  1.2& -1.9&  -6.2\\
   7 e&    6.082&$\pm$  8.8& 14.7&-14.3& 13.1&-12.7&  6.7& -6.5&$\pm$  3.0&  1.2&  2.4&  5.1& -4.9&  -5.8\\
\hline
   8 a&    6.077&$\pm$  9.4& 15.0&-12.6& 13.6&-11.6&  6.3& -5.0&$\pm$  2.3&  0.6&  2.0&  5.2& -3.5&  -6.2\\
   8 b&    6.759&$\pm$  8.5& 13.3&-11.9& 11.7&-10.5&  6.3& -5.6&$\pm$  3.8&  2.3&  1.9&  3.8& -2.4&  -5.6\\
   8 c&    8.305&$\pm$  8.0& 11.0&-10.2& 10.2& -9.5&  4.1& -3.5&$\pm$  0.6&  0.9&  1.0&  3.6& -2.8&  -4.8\\
   8 d&    7.520&$\pm$  8.2& 10.2&-11.2&  9.6&-10.4&  3.3& -4.1&$\pm$  0.4&  1.3&  0.3&  2.7& -3.5&  -5.7\\
\hline
\end{tabular}
\rm \normalsize
\caption{Results of the dijet cross section measurement at high $Q^2$ 
 using the inclusive $k_\perp$ jet algorithm.
The presentation is as in table~\ref{tab:res_incl_ik}.
 \label{tab:res_dij_ik_xi}}
\end{table}

\clearpage

\section*{Results of the QCD Analysis}

\subsubsection*{\boldmath $E_T$ Dependence of  $\alps$}

\begin{table}[h!] \footnotesize \sf  \centering
\begin{tabular}{|l|r|r|r|r|}
\hline
\multicolumn{5}{|c|}{$E_T$ dependence of $\alpset$ \hskip9mm ($\mu_r=E_T$)} \\
\multicolumn{5}{|c|}{inclusive jet cross section --- inclusive $\kperp$ algorithm} \\
\hline
\hline
average $E_T$ of data point:   & $\sqrt{70}\GeV$  & $\sqrt{200}\GeV$ & $\sqrt{500}\GeV
$&
$\sqrt{1500}\GeV$ \\
\hline
\hline
$\alpset=$ &$0.1940$ &$0.1636$ &$0.1579$ & $0.1507$ \\
\hline
total uncertainty &$^{+0.0157}_{-0.0145}$ &$^{+0.0145}_{-0.0136}$ &
   $^{+0.0162}_{-0.0167}$ &$^{+0.0308}_{-0.0315}$  \\ 
\hline
 \,\,\,exp. &$^{+0.0082}_{-0.0081}$ &$^{+0.0106}_{-0.0104}$ &
      $^{+0.0124}_{-0.0123}$ &$^{+0.0264}_{-0.0244}$  \\ 
 \,\,\,theor. &$^{+0.0097}_{-0.0103}$ &$^{+0.0077}_{-0.0081}$ &
         $^{+0.0091}_{-0.0086}$ &$^{+0.0113}_{-0.0088}$  \\
 \,\,\,pdf &$^{+0.0092}_{-0.0061}$ &$^{+0.0040}_{-0.0033}$ &
        $^{+0.0051}_{-0.0072}$ &$^{+0.0110}_{-0.0179}$  \\ 
\hline 
\hline
 $\alpsmz=$ & $0.1211$ & $0.1174$ & $0.1227$ & $0.1292$  \\
\hline
total uncertainty &$^{+0.0058}_{-0.0056}$ &$^{+0.0068}_{-0.0070}$ &
         $^{+0.0094}_{-0.0101}$ &$^{+0.0218}_{-0.0233}$  \\ 
\hline
 \,\,\,exp. &$^{+0.0031}_{-0.0031}$ &$^{+0.0052}_{-0.0053}$ &
        $^{+0.0072}_{-0.0074}$ &$^{+0.0187}_{-0.0182}$  \\ 
 \,\,\,theor. &$^{+0.0035}_{-0.0040}$ &$^{+0.0038}_{-0.0043}$ &
  $^{+0.0053}_{-0.0054}$ & $^{+0.0079}_{-0.0067}$  \\
 \,\,\,pdf &$^{+0.0035}_{-0.0023}$ &$^{+0.0020}_{-0.0017}$ &
 $^{+0.0030}_{-0.0042}$ &$^{+0.0078}_{-0.0129}$  \\
\hline
\end{tabular}
\rm \normalsize
\caption{The $\alps$ results from the fits presented 
in section~\ref{sec:as_et}.
Displayed are the fit results of $\alpset$ at different $E_T$ (top) and 
the corresponding values extrapolated to $\mu_r = M_Z$ (bottom) together 
with the different contributions to the uncertainty.}
\label{tab:as_et}
\end{table}

\subsubsection*{\boldmath The Gluon Density in the Proton}

\begin{table}[h!] \footnotesize \sf  \centering
\begin{tabular}{|r|r||r|r|r|}
\hline
\multicolumn{5}{|c|}{The Gluon Density
  in the Proton at $\mu_f=\sqrt{200}\GeV$} \\
\multicolumn{5}{|c|}{determined for
$\alpsmz = 0.1184 \pm 0.0031$} \\
\multicolumn{5}{|c|}{
parameterized by \,\, $xG(x)\, =\, A x^b \, (1-x)^c \, (1+dx)$ 
\,\, in $0.01<x<0.1$ } \\
\multicolumn{5}{|c|}{central result: \,\,\,
 A=0.503 ;  b = --0.5935 ;  c = 4.70 ;   d = --0.55} \\     
\hline
\hline
\hskip5mm $\log_{10}(x)$ & \hskip14mm $xG(x)=$ &  \hskip7mm exp.\  & 
\hskip7mm theor.\  & from $\Delta \alpsmz$  \\
\hline
-2.0  & $7.35\;^{+1.34}_{-1.25}$ & $^{+0.93}_{-0.94}$& $^{+0.77}_{-0.71}$  & $^{+0.57}_{-0.43}$ \\
-1.9  & $6.32\;^{+1.04}_{-0.98}$ &$^{+0.64}_{-0.65}$ & $^{+0.65}_{-0.63}$  & $^{+0.49}_{-0.37}$ \\
-1.8  & $5.42\;^{+0.82}_{-0.77}$ &$^{+0.45}_{-0.45}$ & $^{+0.51}_{-0.54}$  & $^{+0.42}_{-0.32}$ \\
-1.7  & $4.63\;^{+0.65}_{-0.63}$ &$^{+0.32}_{-0.32}$ & $^{+0.44}_{-0.46}$  & $^{+0.36}_{-0.28}$ \\
-1.6  & $3.93\;^{+0.54}_{-0.52}$ &$^{+0.26}_{-0.25}$ & $^{+0.36}_{-0.38}$  & $^{+0.30}_{-0.24}$ \\
-1.5  & $3.31\;^{+0.45}_{-0.43}$ &$^{+0.23}_{-0.21}$ & $^{+0.28}_{-0.31}$  & $^{+0.25}_{-0.20}$ \\
-1.4  & $2.76\;^{+0.37}_{-0.36}$ &$^{+0.21}_{-0.20}$ & $^{+0.22}_{-0.25}$  & $^{+0.21}_{-0.17}$ \\
-1.3  & $2.27\;^{+0.31}_{-0.30}$ &$^{+0.19}_{-0.18}$ & $^{+0.18}_{-0.19}$  & $^{+0.17}_{-0.14}$ \\
-1.2  & $1.84\;^{+0.26}_{-0.25}$ &$^{+0.17}_{-0.16}$ & $^{+0.13}_{-0.14}$  & $^{+0.14}_{-0.12}$ \\
-1.1  & $1.47\;^{+0.22}_{-0.20}$ &$^{+0.15}_{-0.14}$ & $^{+0.12}_{-0.11}$  & $^{+0.11}_{-0.09}$ \\
-1.0  & $1.14\;^{+0.19}_{-0.17}$ &$^{+0.13}_{-0.13}$ & $^{+0.08}_{-0.10}$  & $^{+0.08}_{-0.07}$ \\
\hline
\end{tabular}
\rm \normalsize
\caption{The gluon density in the proton from the fit in 
section~\ref{sec:quarkgluon}.
Displayed are the central results and the total uncertainties 
of the gluon density at eleven values of $x$ in the interval 
$0.01 \le x \le 0.1$ together with the different contributions
to the uncertainty.
Also shown are the parameters $A,b,c,d$ of the central result.}
\label{tab:gluon}
\end{table}

\end{appendix}

\end{document}

%% file: h1auts.tex

C.~Adloff$^{33}$,              
V.~Andreev$^{24}$,             
B.~Andrieu$^{27}$,             
T.~Anthonis$^{4}$,             
V.~Arkadov$^{35}$,             
A.~Astvatsatourov$^{35}$,      
I.~Ayyaz$^{28}$,               
A.~Babaev$^{23}$,              
J.~B\"ahr$^{35}$,              
P.~Baranov$^{24}$
% ...

\bigskip{\it
% ...
$ ^{4}$ Inter-University Institute for High Energies ULB-VUB, Brussels;
          Universitaire Instelling Antwerpen, Wilrijk; Belgium$^{ c}$ \\
% ...
$ ^{23}$ Institute for Theoretical and Experimental Physics,
          Moscow, Russia \\
 $ ^{24}$ Lebedev Physical Institute, Moscow, Russia$^{ e,h}$ \\
% ...
$ ^{27}$ LPNHE, Ecole Polytechnique, IN2P3-CNRS, Palaiseau, France \\
 $ ^{28}$ LPNHE, Universit\'{e}s Paris VI and VII, IN2P3-CNRS,
          Paris, France \\
% ...
$ ^{33}$ Fachbereich Physik, Bergische Universit\"at Gesamthochschule
          Wuppertal, Wuppertal, Germany$^{ a}$ \\
% ...
$ ^{35}$ DESY, Zeuthen, Germany$^{ a}$ \\
% ...
$ ^a$ Supported by the Bundesministerium f\"ur Bildung, Wissenschaft,
      Forschung und Technologie, FRG,
      under contract numbers 7AC17P, 7AC47P, 7DO55P, 7HH17I, 7HH27P,
      7HD17P, 7HD27P, 7KI17I, 6MP17I and 7WT87P \\
% ...
$ ^c$ Supported by FNRS-NFWO, IISN-IIKW \\
% ...
$ ^e$ Supported by the Deutsche Forschungsgemeinschaft \\
% ...
$ ^h$ Supported by Russian Foundation for Basic Researc
      grant no. 96-02-00019 \\
% ...
}

%% file: paper.bbl
\begin{thebibliography}{10}

\bibitem{heraqcd}
{H1 Collaboration, S. Aid et al.}{,}
\newblock Nucl. Phys. B 449 (1995) 3; \\
ZEUS Collaboration, M. Derrick et al., Phys. Lett. B 363 (1995) 201; \\
H1 Collaboration, C. Adloff et al., Eur. Phys. J. C5 (1998) 625; \\
H1 Collaboration, C. Adloff et al., Eur. Phys. J. C6 (1999) 575.

\bibitem{webber93}
{B.R. Webber}{,}
\newblock J. Phys. G19 (1993) 1567.

\bibitem{ktdis}
{S. Catani, Yu.L. Dokshitzer and B.R. Webber}{,}
\newblock Phys. Lett. B 285 (1992) 291.

\bibitem{cambridge}
{Yu.L. Dokshitzer, G.D. Leder, S. Moretti and B.R. Webber}{,}
\newblock JHEP 08 (1997) 001.

\bibitem{wobisch2000a}
{M.~Wobisch}{,}
\newblock PhD Thesis, RWTH Aachen (2000) PITHA 00/12.

\bibitem{ellissoper}
{S.D. Ellis and D.E. Soper}{,}
\newblock Phys. Rev. D 48 (1993) 3160.

\bibitem{ktclus}
{S. Catani, Yu.L. Dokshitzer, M.H. Seymour and B.R. Webber}{,}
\newblock Nucl. Phys. B 406 (1993) 187.

\bibitem{wobisch99b}
{M.~Wobisch and T.~Wengler}{,}
\newblock Proceedings of the HERA Monte Carlo Workshop, eds. G.~Grindhammer,
  G.~Ingelman, H.~Jung, T.~Doyle, DESY-PROC-02-1999 (1999) 270.

\bibitem{snowmass}
{J. Huth et al.}{,}
\newblock Proceedings of the Summer Study on High Energy Physics, Snowmass,
  Colorado (1990) 134.

\bibitem{frixione97}
{S. Frixione and G. Ridolfi}{,}
\newblock Nucl. Phys. B 507 (1997) 315.

\bibitem{disent}
{ S. Catani and M.H. Seymour}{,}
\newblock Nucl.\ Phys. B 485 (1997) 291, Erratum-ibid. B 510 (1997) 503.

\bibitem{wobisch99c}
{C.~Duprel, Th.~Hadig, N.~Kauer and M.~Wobisch}{,}
\newblock Proceedings of the HERA Monte Carlo Workshop, eds. G. Grindhammer, G.
  Ingelman, H. Jung, T. Doyle, DESY-PROC-02-1999 (1999) 142.

\bibitem{disaster}
{D. Graudenz}{,}
\newblock DISASTER++: Version 1.0 (October 1997), hep-ph/9710244.

\bibitem{HERWIG}
{G. Marchesini et al.}{,}
\newblock Comp.\ Phys.\ Comm.\ 67 (1992) 465.

\bibitem{LEPTO}
{ G. Ingelman, A. Edin and J. Rathsman}{,}
\newblock Comp.\ Phys.\ Comm.\ 101 (1997) 108.

\bibitem{RAPGAP}
{H.~Jung}{,}
\newblock Comp.\ Phys.\ Comm.\ 86 (1995) 147.

\bibitem{ARIADNE}
{L. L\"onnblad}{,}
\newblock Comp.\ Phys.\ Comm.\ 71 (1992) 15.

\bibitem{cluster}
{B.R. Webber}{,}
\newblock Nucl. Phys. B 238 (1984) 492.

\bibitem{lund}
{B. Andersson et al.}{,}
\newblock Phys. Rep. 97 (1983) 31.

\bibitem{heracles}
{A. Kwiatkowski, H. Spiesberger and H.-J. M\"ohring}{,}
\newblock Comp.\ Phys.\ Comm.\ 69 (1992) 155.

\bibitem{django}
{K.\ Charchula, G.\ Schuler and H.\ Spiesberger}{,}
\newblock Comp.\ Phys.\ Comm.\ 81 (1994) 381.

\bibitem{carli99}
{T.~Carli}{,}
\newblock Proceedings of the HERA Monte Carlo Workshop, eds. G.~Grindhammer,
  G.~Ingelman, H.~Jung, T.~Doyle, DESY-PROC-02-1999 (1999) 185.

\bibitem{H1det}
{H1 Collaboration, I. Abt et al.}{,}
\newblock Nucl.\ Instr.\ Meth.\ A 386 (1997) 310 and 348.

\bibitem{h1lar}
{H1 Calorimeter Group, B.~Andrieu et al.}{,}
\newblock Nucl. Instr. Meth. A 336 (1993) 460.

\bibitem{h1lar2}
{H1 Calorimeter Group, B.~Andrieu et al.}{,}
\newblock Nucl. Instr. Meth. A 350 (1994) 57.

\bibitem{h1pions}
{H1 Calorimeter Group, B. Andrieu et al.}{,}
\newblock Nucl. Instr. Meth. A 336 (1993) 499.

\bibitem{h1spacal}
{H1 Spacal Group, R.D.~Appuhn et al.}{,}
\newblock Nucl. Instr. Meth. A 386 (1997) 397.

\bibitem{h1highq2}
{H1 Collaboration, C. Adloff et al.}{,}
\newblock Eur.\ Phys.\ J.\ C 13 (2000) 609.

\bibitem{elecsigma}
{U.~Bassler and G.~Bernardi}{,}
\newblock Nucl. Instr. Meth. A 426 (1999) 583.

\bibitem{pythia}
{T. Sj\"ostrand and M. Bengtsson}{,}
\newblock Comp.\ Phys.\ Comm.\ 43 (1987) 74.

\bibitem{phojet}
{R. Engel}{,}
\newblock Z. Phys. C 66 (1995) 203;\\
R. Engel and J. Ranft, Phys. Rev. D 54
  (1996) 4244.

\bibitem{geant}
{CERN}{,}
\newblock GEANT, detector description and simulation tool, CERN Program Library
  Long Writeup W 5013 (1994).

\bibitem{h1jetstr}
{H1 Collaboration, C. Adloff et al.}{,}
\newblock Nucl. Phys. B 545 (1999) 3.

\bibitem{cteq5}
{H.L. Lai et al.}{,}
\newblock Eur.\ Phys.\ J.\ C 12 (2000) 375.

\bibitem{h1fwdjet99}
{H1 Collaboration, C. Adloff et al.}{,}
\newblock Nucl.\ Phys. B 538 (1999) 3.

\bibitem{zijlstra92}
{E.B. Zijlstra and W.L. van Neerven}{,}
\newblock Nucl. Phys. B 383 (1992) 525.

\bibitem{mepjet}
{E.~Mirkes and D.~Zeppenfeld}{,}
\newblock Phys.~Lett. B 380 (1996) 205.

\bibitem{minuit}
{F. James and M. Roos}{,}
\newblock Comp.\ Phys.\ Comm.\ 10 (1975) 343.

\bibitem{zomer95}
{C. Pascaud and F. Zomer}{,}
\newblock preprint LAL 95-05 (1995).

\bibitem{botje99}
{M.~Botje}{,}
\newblock Eur.\ Phys.\ J.\ C 14 (2000) 285.

\bibitem{zomer99}
{V. Barone, C. Pascaud and F. Zomer}{,}
\newblock Eur.\ Phys.\ J.\ C 12 (2000) 243.

\bibitem{h1compositeness99}
{H1 Collaboration, C. Adloff et al.}{,}
\newblock Phys.\ Lett.\ B 479 (2000) 358.

\bibitem{4loop}
{T. van Ritbergen, J.A.M. Vermaseren and S.A. Larin}{,}
\newblock Phys. Lett. B 400 (1997) 379.

\bibitem{4loopmatch}
{K.G. Chetyrkin, B. A. Kniehl and M. Steinhauser}{,}
\newblock Phys. Rev. Lett. 79 (1997) 2184.

\bibitem{bethke99}
{S. Bethke}{,}
\newblock J. Phys. G 26 (2000) R27.

\bibitem{mrst99}
{A.D. Martin, R.G. Roberts, W.J. Stirling and R.S. Thorne}{,}
\newblock Eur.\ Phys.\ J.\ C 14 (2000) 133.

\bibitem{cteq4}
{CTEQ Collaboration, H.L. Lai et al.}{,}
\newblock Phys.\ Rev.\ D 55 (1997) 1280.

\bibitem{mrsr}
{A.D.~Martin, R.G.~Roberts and W.J.~Stirling}{,}
\newblock Phys. Lett. B 387 (1996) 419.

\bibitem{mrsap}
{A.D.~Martin, W.J.~Stirling and R.G.~Roberts}{,}
\newblock Phys. Lett. B 356 (1995) 89.

\bibitem{cteqgluon}
{J. Huston et al.}{,}
\newblock Phys.\ Rev.\ D 58 (1998) 114034.

\bibitem{grv98}
{M. Gl\"uck, E. Reya and A. Vogt}{,}
\newblock Eur. Phys. J. C 5 (1998) 461.

\bibitem{jadeopal}
{JADE Collaboration and OPAL Collaboration, P. Pfeifenschneider et al.}{,}
\newblock CERN-EP-99-175, to appear in Eur.\ Phys.\ J.\ C.

\end{thebibliography}
